\documentstyle[emulateapj,psfig,apjfonts,astron_mlb]{article}
\submitted{Received July 11 2001}
\newcommand{\shortcite}{\cite*}
\def\etal{{ et al.\thinspace}}

\def\gtrsim{\mathrel{\raise0.35ex\hbox{$\scriptstyle >$}\kern-0.6em
\lower0.40ex\hbox{{$\scriptstyle \sim$}}}}
\def\lesssim{\mathrel{\raise0.35ex\hbox{$\scriptstyle <$}\kern-0.6em
\lower0.40ex\hbox{{$\scriptstyle \sim$}}}}

\def\mtot{\hbox{$\rm\thinspace M_{702}$}}

\def\lowlx{Low--$L_X$}
\def\hilx{High--$L_X$}
\def\h50{h_{50}}
\def\MDR{T--$\Sigma$}

\begin{document} 

\title{Distinguishing Local and Global Influences on Galaxy Morphology: \\
An {\it HST} Comparison of High and Low
X-ray Luminosity Clusters\footnotemark}

\lefthead{Balogh et al.}
\righthead{The Environmental Dependence of Galaxy Morphology}

\author
{Michael L.\ Balogh,$\!$\altaffilmark{2,3} Ian Smail,$\!$\altaffilmark{2} R.\,G.\ Bower,$\!$\altaffilmark{2} B.\,L.\ Ziegler,$\!$\altaffilmark{4} 
G.\,P.\ Smith,$\!$\altaffilmark{2} Roger L.\ Davies,$\!$\altaffilmark{2}
\\ A.\ Gaztelu,$\!$\altaffilmark{2} 
J.-P.\ Kneib$^{5}$ \& H.\ Ebeling$^{6}$\\ 
}

\setcounter{footnote}{1}

\footnotetext{Based on observations with the NASA/ESA {\it Hubble Space
Telescope} obtained at the Space Telescope Science Institute, which is
operated by the Association of Universities for Research in Astronomy
Inc., under NASA contract NAS 5-26555.}

\altaffiltext{2}{Department of Physics, University of Durham,
South Road,
Durham DH1 3LE, UK}
\altaffiltext{3}{E-mail: m.l.balogh@durham.ac.uk}
\altaffiltext{4}{Universitaetssternwarte, Geismarlandstr. 11, 37083 Goettingen, Germany}
\altaffiltext{5}{Observatoire Midi-Pyr\'en\'ees, CNRS-UMR5572,
14 Avenue E.\,Belin, 31400 Toulouse, France}
\altaffiltext{6}{Institute for Astronomy, 2680 Woodlawn Drive, Honolulu, Hawaii 96822 USA}

\setcounter{footnote}{6}

\begin{abstract}
We present a morphological analysis of 17 X-ray selected clusters at $z
\sim0.25$, imaged uniformly with {\it Hubble Space Telescope WFPC2}. Eight
of these clusters comprise a subsample selected for their low X-ray
luminosities ($\lesssim 10^{44}$\,erg\,s$^{-1}$), called the \lowlx\
sample. The remaining nine clusters comprise a \hilx\ subsample with
$L_X> 10^{45}$\,ergs\,s$^{-1}$.  The two subsamples differ in
their mean X-ray luminosity by a factor of 30, and span a range of
more than 300.  The clusters cover a relatively small range in redshift
($z=$0.17--0.3, $\sigma_z/z\sim0.15$) and the data are homogeneous in terms of depth, resolution
(0\farcs17$=1\h50^{-1}$ kpc at $z=0.25$) and rest wavelength observed,
minimizing differential corrections from cluster to cluster.  We fit the
two dimensional surface brightness profiles of galaxies down to very
faint absolute magnitudes: $\mtot\leq -18.2+5\log{\h50}$ (roughly
$0.01 L^\ast_R$) with parametric models, and quantify their morphologies using the
fractional bulge luminosity (B/T).   Within a single {\it WFPC2} image,
covering a field of $\sim 3\arcmin$ ($1 \h50^{-1}$\,Mpc at $z=0.25$)
in the cluster centre, we find that the \lowlx\ clusters are dominated by
galaxies with low B/T ($\sim 0$), while the \hilx\ clusters are dominated
by galaxies with intermediate B/T ($\sim 0.4$).
We test whether this difference could arise
from a universal morphology-density relation due to differences in
the typical galaxy densities in the two samples.  We find that small
differences in the B/T distributions of the two samples persist with
marginal statistical significance (98\% confidence based on a binned $\chi^2$ test) even
when we restrict the comparison to galaxies in environments with similar
projected local galaxy densities.  A related
difference (also of low statistical significance) is seen between the bulge
luminosity functions of the two cluster samples, while no difference is
seen between the disk luminosity functions.  From the correlations 
between these quantities, we argue that   
the global environment affects the population of bulges, over and
above trends seen with local density.  On the basis
of this result we conclude that the destruction of disks through
ram pressure stripping or harassment is not solely responsible for the
morphology-density relation, and that bulge formation is less efficient
in low mass clusters, perhaps reflecting a less rich merger history.
\end{abstract}

\section{Introduction}

The local and large-scale environment is known to be an important
determinant of many galaxy properties, in particular star formation rates
(\cite{O60,D+85,B+98,MW00}),
gas content (\cite{B77,GH85,WS91,Bravo+00,V+01,Solanes01})
and especially morphology (\cite{HH31,SB51,M61,A65,Dressler,PG84}).
These studies all suggest that galaxy morphology, in particular
the presence of a star-forming disk, is a
fundamental defining characteristic of galaxies (Hubble \shortcite{Hubble1};\shortcite{Hubbleseq}).

The variation in morphological mix with local galaxy density was first
quantified by the work of Dressler (\shortcite{Dressler}), who showed that the proportion
of early-type (elliptical and lenticular) galaxies increases at the
expense of late-type spiral galaxies in regions with higher projected
galaxy density.  This trend is called the morphology-density relation,
hereafter \MDR, and in this paper we investigate whether the \MDR\ can
provide any insight into the origin and permanence of galaxy morphology.

The advent of high-resolution imaging from the {\it
Hubble Space Telescope} ({\it HST}) has allowed greatly improved
quantitative analyses of galaxy morphologies, especially beyond the very local universe
($z\gtrsim0.1$).  This has resulted in a renewed interest in the use
of morphology as an important tracer of galaxy evolution (e.g. Couch \etal\ \shortcite{C+94}; Abraham \etal\ \shortcite{A+96};
Smail \etal\ \shortcite{Smail}; Schade, Barrientos \& Lopez-Cruz \shortcite{Schade97}).  
For example, working from an extensive
{\it HST} survey of rich clusters at $z\gtrsim $0.4--0.5, Dressler et al.\
(\shortcite{D+97}) have shown that the \MDR\ seen in local clusters is present in
the most massive systems at $z\sim 0.5$.  However, these studies have
also suggested that a galaxy's morphology may evolve, complicating the
comparison of populations at different epochs.
  
Theoretically, there are several physical mechanisms which may
determine or alter a galaxy's morphology as a function of local density
and/or cosmic time.  Early numerical simulations (\cite{RN,FS82,Barnes})
demonstrated that a merger
between two spiral galaxies would almost always result in a galaxy with
structural parameters resembling those of elliptical galaxies.  Thus,
environments rich in merger events (dense regions with low velocity
dispersion) are expected to be fertile regions for the formation of
elliptical galaxies.  This is an important ingredient in modern models
of galaxy formation (\cite{Cole91,K96,KWG,semianal,SP99})
which
produce a reasonably good match to the observed \MDR\  (\cite{Diaferio}).
However, there are other physical mechanisms, generally
not included in these models, which might affect galaxy morphology.
Moore et al.\ (\shortcite{harass})
showed that tidal interactions between galaxies in
dense environments are effective at destroying galactic disks and at
transforming spiral galaxies into the dwarf spheroidal galaxies that
dominate rich clusters today.  Other processes, such as ram pressure
stripping (\cite{GG})
turbulent and viscous stripping (\cite{N82})
and  strangulation (\cite{LTC,infall})
result in the fading of galactic disks and in the smoothing out of
their surface brightness due to the fading of H{\sc II} regions.
These processes, then, might transform spiral galaxies into S0 galaxies
or anemic spirals (\cite{vdB91}).

One of the stronger trends seen in the \MDR\ at low redshift is the
dramatic increase in the number of lenticular (S0) galaxies in the
highest density regions of the universe (\cite{Dressler}).
There is
now also evidence that the fraction of S0 galaxies in rich clusters
evolves dramatically with redshift (Dressler et al.\ \shortcite{D+97}; Fasano
et al.\ \shortcite{F+00}, but see also Andreon \shortcite{Andreon98} and Fabricant, Franx \& van
Dokkum \shortcite{FFD}),
suggesting that the dominant process in defining the
\MDR\ is the transformation of spiral galaxies into S0 galaxies within
high density regions (\cite{P+99,KS00}).
This suggestion appears to be consistent with studies of age differences
between elliptical and S0 galaxies in clusters (\cite{KD98,vd98,JSC,Smail01}).
However, there remain significant questions about the feasibility of
this transformation.  Most importantly, it has been suggested that the
correlation between bulge luminosity and bulge-to-disk ratio for spiral
galaxies means that it is difficult to form the entire cluster S0
population by stripping the disks of spirals, as there is too little
luminosity density in spiral bulges (\cite{YW75,Dressler,SSS,KS00}).
Since the observed
evolution in this galaxy population occurs at relatively modest redshifts
($z\lesssim0.5$), this is an issue which can be addressed with present
observational facilities.

Another unanswered question is which physical processes, acting in
which environments, are responsible for establishing the morphological
mix of a galaxy population.  
Dressler (\shortcite{Dressler}) showed that
galaxy morphology is strongly correlated with projected local galaxy
density, suggesting that morphology must change as the local
density changes, irrespective of whether or not the galaxy is in a rich
cluster.  However, this has been difficult to establish with certainty
because, particularly in rich clusters, local galaxy density is closely
correlated with cluster-centric distance.  It has been suggested that,
with an appropriate choice of the cluster centre, morphology correlates
better with radius than with local density and hence this global property
is more likely to be driving the variation (\cite{WG,WGJ}).

To address this question observationally,
it is critical to investigate how galaxy properties depend on global
cluster properties such as mass or X-ray luminosity, independent of any
other factors, such as local density or redshift.  One can then hope to
distinguish between physical mechanisms which affect galaxy properties
on very local scales from those which are coupled to the large-scale
environment.  For example, ram pressure stripping (\cite{GG})
is strongly dependent on the galaxy velocity in the presence of a dense
intracluster medium, and is thus more likely to occur within massive
clusters with high velocity dispersions.   On the other hand, galaxy
merging is most effective in regions of low velocity, so may be more
important in groups and low mass clusters.  Finally, disk galaxies
which end up in massive clusters may have intrinsically larger
bulge-to-disk ratios than the general field population.  This could
occur since it is expected that galaxies in regions which are overdense
on large scales form earlier than average (\cite{BBKS,NBW}).
As a result, such galaxies may
have consumed more of their gas and be more likely to have reduced or
ceased star formation in their disks by the present day; consequently,
the galaxies would be more bulge-dominated.

To address these issues, we have obtained {\it Hubble Space Telescope}
({\it HST})  {\it WFPC2} F702W ($R_{702}$-band) imaging of 17 clusters
at $0.17<z<0.30$ (with a mean redshift of $z=0.25$), selected from two
published {\it ROSAT} cluster samples.   The choice of redshift is a
compromise between the need for both deep and wide images (to ensure
coverage of a large physical radius); furthermore,
$z=0.25$ represents an epoch where a significant S0 galaxy population
is visible in the cores of rich clusters, and must have been very
recently transformed.  Therefore, observations at this epoch represent
the best opportunity of uncovering the dominant environment-related
processes  responsible for the creation of S0 galaxies
(\cite{JSC,Smail01}).

All clusters were selected on the basis of their X-ray
emission.  As part of a more extensive imaging and spectroscopic study
underway, we selected eight  clusters from the sample of  Vikhlinin et
al.\ (\shortcite{V+98})
with the lowest X-ray fluxes, $L_X \leq 1.5 \times 10^{44}
h_{50}^{-2}$ ergs s$^{-1}$, in the redshift range $z=0.2$--0.3.  A
detailed analysis of these eight clusters, including results from
extensive follow-up spectroscopy, which is in preparation, will provide
a crucial link between studies of massive clusters (e.g. Balogh \etal\ \shortcite{PSG}; Poggianti \etal\ \shortcite{P+99})
and groups (e.g. Zabludoff \& Mulchaey \shortcite{ZM98}; Tran \etal\ \shortcite{Tran}). 
Complementing these are nine clusters with
high X-ray luminosities $L_X \geq 10^{45} h_{50}^{-2}$ ergs
s$^{-1}$ at 0.1--2.4 keV, and taken from the XBACS sample of
Ebeling et al.\ (\shortcite{XBACS}).
The mean $L_X$ of the \hilx\ clusters is a
factor of 30 larger than that of the low luminosity cluster sample.  It
is expected that this difference in $L_X$ directly corresponds to a
difference in mass, since the correlation ($L_X\propto M^2$) is well
established observationally and understood theoretically in this 
$L_x$ range (\cite{L+99,Cooray,WXF,XW,Babul2}).
While the two samples have very different
X-ray properties, they cover an overlapping range in local galaxy density.
We can therefore test directly how galaxy properties depend separately
on cluster mass and local density.

%
% Table 1
%
{\scriptsize
\begin{table*} 
\begin{center} 
\caption{\centerline {\sc \label{tab-1} Log of the {\it HST} observations of the cluster samples}} 
\vspace{0.1cm}
\begin{tabular}{lccccccc} 
\hline\hline
\noalign{\smallskip}
Name      &   R.A.\   & Dec.\       & $z$ & Exposure &$L_X$ (0.1--2.4 keV)                 & $R^{\rm lim}_{702}$& $N>R^{\rm lim}_{702}$\cr 
          & \multispan2{\hfil (J2000)\hfil }   &          & time (ks) &$10^{44} \h50^{-2}$ ergs s$^{-1}$ &       &                \cr 
\noalign{\smallskip}
\hline
\noalign{\smallskip}
\multispan2{Low--$L_X$~Sample\hfil }   &       &     &      &     &           &    \cr                                    \noalign{\smallskip}
Cl\,0818+56 & 08 18 58 & +56 54 34 & 0.248    & ~7.2     & 0.56                               & 23.00        & ~86 \cr
Cl\,0819+70 & 08 19 23 & +70 54 48 & 0.230    & ~6.9     & 0.48                               & 22.80        & ~34 \cr
Cl\,0841+70 & 08 41 43 & +70 46 53 & 0.240    & ~6.9     & 0.46                               & 22.90        & ~39 \cr
Cl\,0849+37 & 08 49 11 & +37 31 25 & 0.235    & ~7.8     & 0.73                               & 22.85        & ~68 \cr
Cl\,1309+32 & 13 09 56 & +32 22 31 & 0.294    & ~7.8     & 0.74                               & 23.40        & ~99 \cr
Cl\,1444+63 & 14 44 08 & +63 44 58 & 0.297    & ~7.5     & 1.46                               & 23.45        & ~96 \cr
Cl\,1701+64 & 17 01 46 & +64 21 15 & 0.222    & ~7.5     & 0.15                               & 22.70        & ~64 \cr
Cl\,1702+64 & 17 02 13 & +64 20 00 & 0.242    & ~7.5     & 0.33                               & 22.90        & ~61 \cr
\noalign{\smallskip}
\hline
\noalign{\smallskip}
\multispan2{High--$L_X$~Sample\hfil }     &       &     &      &     &           &    \cr
\noalign{\smallskip}
A\,68       & 00 36 59    & +09 08 30  & 0.255    & ~7.5  & 18.7                               & 23.00        &102 \cr
A\,267      & 01 52 52    & +01 02 46  & 0.230    & ~7.5  & 16.9                               & 22.80        & ~90 \cr
A\,383      & 02 48 07    & $-$03 29 32  & 0.185  & ~7.5  & 11.7                               & 22.30        & ~73 \cr
A\,773      & 09 17 59    & +51 42 23  & 0.217    & ~7.2  & 16.0                               & 22.90        &155 \cr
A\,963      & 10 17 10    & +39 01 00  & 0.206    & ~7.8  & 12.6                               & 22.50        & ~96 \cr 
A\,1763     & 13 35 17    & +40 59 58  & 0.228    & ~7.8  & 11.8                               & 22.80        & ~116 \cr
A\,1835     & 14 01 02    & +02 51 32  & 0.253    & ~7.5  & 48.3                               & 23.00        &141 \cr
A\,2218     & 16 35 54    & +66 13 00  & 0.171    & ~6.5  & 10.9                               & 22.10        & ~97 \cr
A\,2219     & 16 40 21    & +46 41 16  & 0.228    & ~14.4 & 25.1                               & 22.77        &110 \cr
\noalign{\hrule}
\end{tabular}
\end{center} 
\end{table*}
}

To enable the cleanest comparison between the two samples we have
ensured that  the observational data collected are as homogeneous
as possible in terms of depth, filter and detector
characteristics. Moreover, since the clusters are all at similar
redshifts, there is minimal variation in k-corrections, spatial
resolution, background contamination, field of view, and absolute
magnitude limit between them.

The data samples and measurements are described in \S2.  In \S3 we
show the dependence of the morphological mix of the galaxies on
cluster X-ray luminosity and projected local galaxy density.   
We discuss the observed correlations in the context of models in which
the disks or bulges of galaxies alone are altered, in \S4.
Our findings are summarized in \S5.  For all cosmology-dependent
calculations we assume $\Lambda=0.7$, $\Omega_m=0.3$ ($\Lambda$CDM)
and parametrise the Hubble constant as $H_\circ = 50 \h50$ km s$^{-1}$
Mpc$^{-1}$.

\section{Observations}\label{sec-obs}

\subsection{Data}\label{sec-data}

\subsubsection{\lowlx\ Sample}  

We selected eight X-ray faint clusters in the northern hemisphere clusters  from the sample identified by Vikhlinin et al.\ (\shortcite{V+98})
in serendipitous, pointed {\it ROSAT} {\it PSPC} observations.
The sample was restricted to a
relatively narrow redshift range, $z=0.22$--0.29 ($\sigma z/z\sim 0.1$)
and a mean redshift of $z=0.25$, to reduce the effects of differential
distance modulus and k-correction effects on the comparison between
the systems.  The X-ray luminosities of these systems range from
$0.15$ to $1.5\times 10^{44} \h50^{-2}$ ergs s$^{-1}$[0.1--2.4 keV] (Table~1). 
We compute $L_X$ in the 0.1--2.4 keV band from the observed
fluxes in the 0.5--2.0 keV band, corrected for galactic H{\sc i} absorption and assuming a k-correction
appropriate for an intra-cluster gas temperature equal to that expected from
the local $L_x-kT$ relation (\cite{AF,Markevitch}),
using the software package {\sc xspec}.  We adopt cluster redshifts obtained 
in the course of our own spectroscopic follow-up observations, the results of
which will be published separately; our redshifts
agree well with the redshifts published in Vikhlinin et al.\ (\shortcite{V+98}).
We refer to these clusters as the \lowlx\ sample.  

The properties of the \lowlx\ clusters
are summarized in Table~1, where we list the cluster name (column 1),
J2000 coordinates (2,3), mean redshift from our spectroscopic data (4),
exposure time in ks (5) and $L_X$ [0.1--2.4 keV] 
in units of $\h50^{-2}$ ergs s$^{-1}$ (6).  
We also list in column 7 the magnitude limit adopted
in our analysis (see \S2.3), and in column 8 the number of
galaxies brighter than this limit.

For each cluster, three single orbit exposures with {\it WFPC2} in the
F702W filter were obtained with {\it HST} during Cycle~8, with exposure
times ranging from 2100\,s to 2600\,s per orbit.  The pointing positions
for the three exposures were offset by 10 {\it Wide Field Camera  (WFC)}
pixels from one another; during the reduction procedure they were aligned
and coadded to remove cosmic rays and hot pixels.  The images were not
drizzled or regridded, as this does not accurately preserve the noise
characteristics of the pixels, and the surface brightness fitting software
that we use is sensitive to the regridding pattern.  
After coadding the frames, the image from each {\it WFC}  chip was trimmed to the area of full
sensitivity. The three {\it WFC} chips were not mosaicked, and we have
not considered the {\it Planetary Camera} images in this analysis.
The photometry is calibrated on the Vega system, with updated zero
points taken from the current instrument manual.  The final images
reach a 3-$\sigma$ point source sensitivity of $R_{702} \sim 25.5$, and
cover a field of $2.5' \times 2.5'$ (or $0.8\h50^{-1}$\,Mpc at $z=0.25$)
with an angular resolution of 0.17$''$ ($\sim$1$\h50^{-1}$\,kpc).

\subsubsection{\hilx\ Sample} 
The comparison sample of very X-ray luminous clusters
comprises nine clusters with $L_x \geq 10^{45} h_{50}^{-2}$ erg
s$^{-1}$ (0.1--2.4 keV) taken from the XBACS sample of Ebeling \etal\
\shortcite{XBACS}. With redshifts $0.17<z<0.26$ (mean of $z=0.21$)
they lie in a narrow redshift range very comparable to that of the
\lowlx\ sample.  Redshifts and X-ray luminosities are taken from
Ebeling \etal\ \shortcite{Ebeling98} (except the luminosity of A383,
taken from Smith \etal\ \shortcite{GPS1}, and the redshift for Abell
1763, taken from Struble \& Rood 1999\nocite{SR99}).
The cluster specifics are given in Table \ref{tab-1}.

Our \hilx\ clusters are part of a larger sample of 12 for which WFPC2
images were obtained (including some archival data) in the course of
an {\it HST} imaging program focused primarily on gravitational
lensing by intermediate redshift clusters (see Smith \etal\
2001a\nocite{GPS1}, 2001b\nocite{GPS2} for first results). The
parameters of the {\it HST} imaging are the same as detailed before
for the \lowlx\ sample: each cluster is observed with one {\it HST
WFPC} pointing in the F702W filter for three orbits, with the pointing
for each orbit shifted by 10 {\it WFC} pixels to allow the removal of
cosmic rays and hot pixels (with the exception of the archival data
for A2219, which is observed for 6 orbits, and A2218, which is shifted
only 3 pixels between exposures).  The reduction procedure is
identical to that used for the \lowlx\ sample.

\subsubsection{Field Sample}

To provide a reference field sample to compare to the clusters, and
equally importantly to allow us to correct the clusters for fore- and
background contamination, we have also analysed images for eight deep
{\it WFPC2} fields selected from the Medium Deep Survey\footnote{The
Medium Deep Survey catalog is based on observations with the NASA/ESA
Hubble Space Telescope, obtained at the Space Telescope Science
Institute, which is operated by the Association of Universities for
Research in Astronomy, Inc., under NASA contract NAS5-26555. The Medium
Deep Survey is funded by STScI grant GO2684.} 
(MDS, Ostrander \etal\ \shortcite{MDS-data}; Ratnatunga \etal\ \shortcite{MDS-morph3}).  
These images were
obtained from the MDS archive at STScI and are already reduced, but
coadded without rejection of hot pixels.  This degrades the cosmetic quality and the number of useful
pixels, but does not otherwise affect our analysis because bad pixels are flagged and
rejected from the surface brightness fits.

%
% Table 2
%
\begin{center}  
{\scriptsize 
\centerline{\sc Table 2}
\centerline{\sc Log of MDS fields.}
\vspace{0.1cm}
\begin{tabular}{lcc} 
\hline\hline
\noalign{\smallskip}
Field      &  Exposure & $N_{\rm gal}(I_{814}<22.6)$\cr 
          &  time (ks)       &     \cr 
\noalign{\smallskip}
\hline
\noalign{\smallskip}
u2h91     &  28.8           & 34  \cr
uci10     &  10.8            & 27  \cr
umd07     &  ~9.6         & 20  \cr
umd0a     &  ~8.7            & 34  \cr
umd0e     &  ~8.4           & 23  \cr
umd0h     &  ~8.2       & 28  \cr
ust00     &  12.1         & 41  \cr 
uwp00     &  ~8.4         & 24  \cr
\noalign{\hrule}
\end{tabular}
}
\end{center}
\medskip

The MDS images were taken with the F606W and F814W ($I_{814}$) filters.  
We choose to analyse the F814W images, rather
than the F606W images, as the k-correction to F702W (used for the cluster
observations) from F814W is typically
more uniform over the redshift range of interest ($z\lesssim0.8$), compared
with that from F606W, for which the 4000\AA\ break moves into the F606W
passband at $z\sim0.2$.
To estimate the field contamination in our
cluster samples we must calculate the surface density of field
galaxies to a fixed $R_{702}$ magnitude limit.   From the deep
number counts of Metcalfe \etal\ (\shortcite{WHDF5}), the median color
of galaxies brighter than $R_{702}=23$ is $(R_{702}-I_{814})=0.4$, adopting the
transformation $I_{814}\approx I$ and $R_{702}=R-0.2$ (\cite{F+95}).
This is consistent with the median galaxy color $(V-I)\approx1.1$
in the CFRS (\cite{CFRS5}) and the MDS (\cite{MDS-numcts}), at
$I_{814}=22.5$.  The color
distribution of galaxies at this depth is broadly distributed, with
an approximate standard deviation of $0.2$ mag; none of the results in 
this paper are sensitive to variations in the $(R_{702}-I_{814})$ color
within this range.  We find a field galaxy
density of 19440 deg$^{-2}$ brighter than $I_{814}=22.5$ in the MDS
images, fully consistent with the number counts of Metcalfe \etal\ (\shortcite{WHDF5}).
We neglect any
correction in the field density due to lensing by the cluster, which we
expect to be below the level of field-to-field variance  (\cite{GPS2}).

There are several possible biases resulting from adopting F814W observations
for our field correction.  
The first is that galaxies have smoother
profiles and more dominant bulge components in the redder filter.
However,  this is not expected to be a large effect (e.g. Saglia \etal\ \shortcite{S+00}), 
and we do not correct for it.  Also, selection in redder
filters tends to preferentially select bulge-dominated galaxies; however,
the results of a morphological analysis of the MDS in the $V_{606}$ and $I_{814}$
filters (\cite{MDS-data}) show this differential effect to be small.
Finally,
the k-corrections for the bulge and disk components will be different
across our cluster sample.  This differential effect is small enough that
it can be safely ignored over the small redshift range of our sample,
but care must be taken when quantitatively comparing these results
with others at different redshifts.  A difference of 0.1 magnitudes
in the k-correction for the bulge, relative to the galaxy as a whole,
will result in a 10 per cent error in the fractional bulge luminosity.

In Table~2 we list the eight MDS fields used, with their exposure times
in column 2, and
the number of galaxies above the fiducial limiting magnitude for morphological
classification ($I_{814}=22.6$, see \S2.4) in column 3.  Across the
eight fields we find a total of 231 galaxies brighter than $I_{814}=22.6$
($R_{702}\sim 23.0$), with a field-to-field standard deviation of 7 galaxies.

\subsection{Source Detection}

Sources were detected in the {\it HST} images using the SExtractor
software v.2.1.6  (\cite{sextractor}).  
For a source to be accepted the signal in at least
10 of its {\it WFC} pixels (0.1 arcsec$^2$) had to be a minimum of
1.5-$\sigma$ above the background.  The faintest sources which are reliably
detected using these criteria have $R_{702}\sim 25$--25.5. Our results
are insensitive to these criteria, as structural parameters
can only be reliably determined for galaxies well above the detection
limit.  Our results do depend, however, on the deblending parameters,
as the surface brightness fits (see \S2.3) use the segmentation image
produced by SExtractor to identify which pixels belong to the galaxy.
In the first pass, we used 32 deblending sub-thresholds, with a minimum
contrast parameter of $7.5\times 10^{-4}$, which deblends well
the majority of galaxies.  However, in some cases this process is too
effecient, and identifies, for example, bright knots in spiral disks as
separate sources.  An even more troublesome problem arises in crowded
regions, when a smaller galaxy is deblended from the flux profile of a
brighter galaxy.  Often SExtractor correctly identifies the separate centroids,
but incorrectly associates a large number of pixels from the bright
galaxy with the fainter.  In total, about 20 per cent of
the sources were improperly deblended on the first trial.   We therefore
repeat our measurements on these sources, after interactively choosing
more appropriate deblending parameters.

\subsection{Morphology Measurements}\label{sec-gim2d}
%
% Figure 1
%
\begin{figure*}
\centerline{\psfig{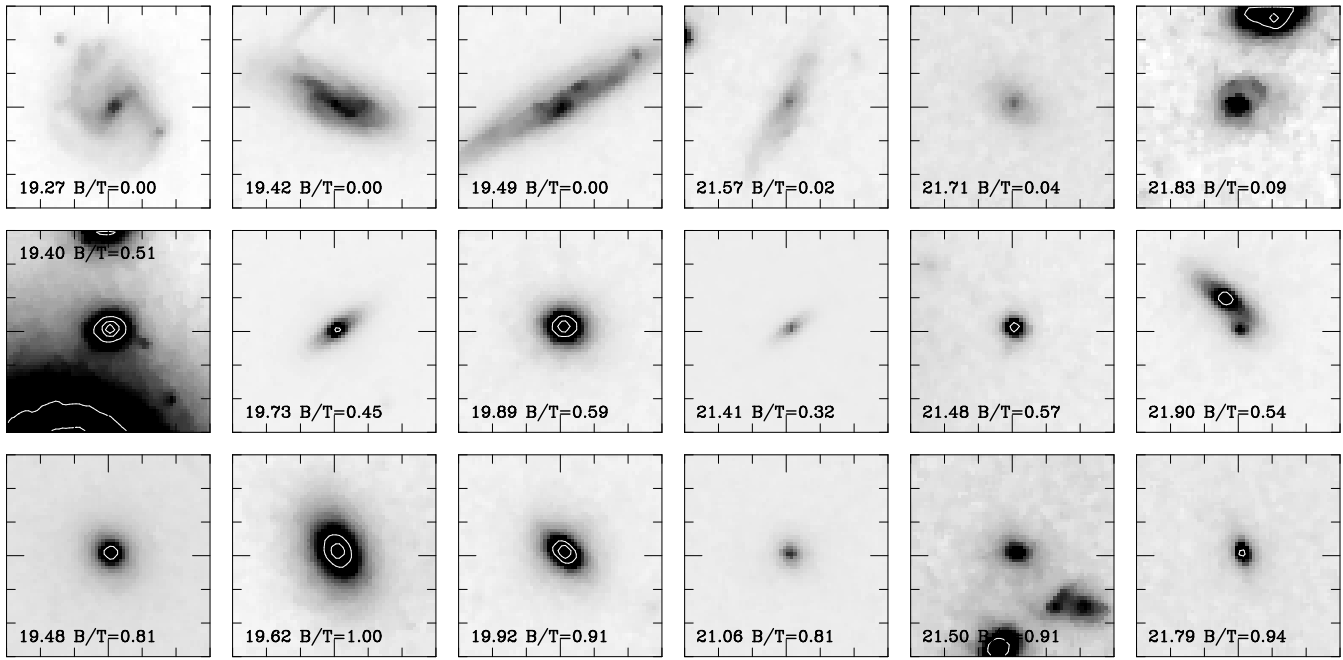}}
\caption{\scriptsize \addtolength{\baselineskip}{-3pt}
Eighteen galaxies (9 bright and 9 faint) randomly selected from the cluster
and field samples.
Apparent $R_{702}$ magnitude and B/T values are shown in each panel.  Each image
is 6\arcsec\ on a side, and contours are
arbitrarily spaced.  The top row shows galaxies with B/T$<0.1$, which are
primarily spirals.  Galaxies in the middle row have $0.4<B/T<0.6$, while
the bottom row shows galaxies with B/T$>0.8$.
}
\end{figure*}

The visual classification of galaxies onto the revised Hubble system (or
variations on it) has remained a central theme of recent morphological
studies of distant galaxies with {\it HST} (\cite{MDS-morph2,Smail,vdB97,C+98}).
However, the high resolution and uniform quality data of these {\it HST} data has
also encouraged the development of automated, machine-based techniques,
which allow quantitative measures of morphology related to, but distinct
from, the revised Hubble system (\cite{Schade95,A+96,O+96,NRG,MS,Brinch98}).
For the purposes of this paper we
have chosen to follow the latter approach and hence analyse our sample
using machine-based classifications based on fits to the two-dimensional
surface brightness profiles of galaxies in the {\it WFPC2} images.

To fit the data we use the {\sc iraf} package {\sc gim2d} v2.2.0, written by Luc
Simard\footnote{http://nenuphar.as.arizona.edu/simard/gim2d/gim2d.html}.
The details of this program are given in Simard et al.\ (\shortcite{Gim2d}).
{\sc gim2d} fits the sky-subtracted surface brightness distribution of
each galaxy with up to twelve parameters describing a ``bulge'' and ``disk'' component.
The algorithm is similar to that used by Schade
et al.\ (\shortcite{S2+96,S3+96}),
except that images are not symmetrized before
fitting.  Clearly, this two component model is a simplified approximation
to the structure of real galaxies; furthermore, the distinction between a
``disk'' and ``bulge'' component is made purely on the surface brightness
profile, and does not necessarily correspond to kinematically distinct
components (see Simard et al.\ \shortcite{Gim2d}).  For simplicity we nevertheless use
the terms disk and bulge for these components in the following discussion.

The disk component is approximated by an exponential model, with parameters
corresponding to the scale length, position angle and inclination.
For the bulge, we fit a simple de Vaucouleurs profile instead of the
more general S\'ersic profile because the former generally provides
a good fit to bright elliptical
galaxies and the bulges of early-type spiral galaxies (\cite{deJong94,Courteau96,A98}).
S\'ersic profiles are more
appropriate for the bulges of late-type galaxies (\cite{deJong94});
however,
these small bulges will be of low signal-to-noise in our images, making an
accurate determination of the S\'ersic index difficult.

{\sc gim2d} searches the large parameter space of models using the
Metropolis (\shortcite{Metropolis}) algorithm, which is inefficient but does not easily get
trapped in local minima.  An important step in this algorithm is an
initial, broad sampling of the parameter space; we populate this space
with 300 models.

We adopt a nominal magnitude limit of $R_{702}=23.0$ at $z=0.25$;
galaxies brighter than this value are large enough and bright enough to be
reliably classified. For $\Lambda$CDM, including small ($\lesssim0.1$
mag) k-corrections, this corresponds to $\mtot=-18.2+5\log{\h50}$,
approximately $M_{R}\sim -18.0+5\log{\h50}$, 4.6 magnitudes below
$M_R^\ast$ (\cite{Sloan_lf}).
Accounting for the range of redshifts spanned by our clusters, 
we adjust this limit so that the
data are matched to the same limiting absolute magnitude. Therefore, for
our most nearby cluster A\,2218 ($z=0.17$) we use a magnitude limit
of $R_{702}=22.1$, while for the most distant cluster Cl\,1444+63
($z=0.29$) we use $R_{702}=23.45$.   We use a field galaxy sample 
cut at the corresponding $I_{814}$ limit based on a typical
color of $(R_{702}-I_{814})\sim 0.4$, as discussed in \S2.1.

We start from an initial list of sources from the SExtractor catalogs
which are more than 3\arcsec\ from a {\it WFC} chip edge, are entirely
contained within the trimmed image, and are brighter than the relevant
magnitude limit for the cluster listed in Table~1.  We then exclude
sources with half-light radii less than 0\farcs15, which are stars, and
giant arcs in the \hilx\ sample.  The exclusion of the arcs, which are
clearly background galaxies, has no significant impact on our results
because the background contamination in the \hilx\ clusters is very small.

For each galaxy, we run {\sc gim2d} using the pixel membership defined
by SExtractor's segmentation image.  This provides best fitting estimates
and errors for the bulge and disk components of the galaxy.  It is quite
remarkable how good a fit this simple, two component model provides to
the majority of the galaxies; we find reduced $\chi^2$ values between one and two
for more than 90 per cent of the galaxies.  No restrictions regarding
goodness of fit or symmetry are imposed in the subsequent analysis.
From the fitted parameters we calculate B/T, the fraction of the luminosity
in the bulge, relative to the total luminosity.  This parameter is
known to correlate with Hubble type (\cite{SdV}),
though there is considerable scatter about the mean relation.
The uncertainties in B/T are determined from Monte Carlo simulation in
the {\sc GIM2D} software, and have a median value of 0.07.

To ensure consistency with the results of the photometric fitting,
the total magnitude for each galaxy is taken to be the total flux in
the model, as computed by
the {\sc GIM2D} software.  These total fluxes are systematically brighter
than the Kron-type magnitudes calculated with SExtractor, by $\sim 0.15$
mag, with a dispersion of $\sim 0.25$ mag.  The offset is larger, $\sim
0.25$ mag, for the brightest galaxies, $R_{702}<18$; fainter than
this magnitude there is
no significant trend with magnitude.

In Figure ~1 we show 6\arcsec\ images of 18 galaxies, randomly selected
from the \lowlx, \hilx\ and MDS samples.  The $R_{702}$ magnitude
and B/T value are printed in each panel.  In the top row, three bright
and three faint galaxies with B/T$<0.1$ are shown.  This regime of B/T consistently
selects disk-dominated galaxies, despite the common occurrence of spiral
arms and bright HII regions.  This is not too surprising, since the
disk decomposition is based on the assumption that the disk surface brightness
follows an exponential profile, as is the case in the majority of spiral
galaxies (e.g. de Jong 1996\nocite{deJong96}).  In the next two rows, we show images of
galaxies with $0.4<B/T<0.6$, and $B/T>0.8$, respectively.  
This allows the reader to associate B/T measurements with the more
traditional morphology of the galaxy (see also Marleau \& Simard et al. 1998\nocite{MS};
Tran et al. 2001\nocite{Tran};  Simard et al. 2001\nocite{Gim2d}).

\section{Results and Analysis}\label{sec-results}

\subsection{Morphological Distributions}\label{sec-mdist}
The luminosity functions
of the \lowlx\ and \hilx\ samples are shown in Figure~2, normalized to the average
number of galaxies per cluster above the magnitude limit in the \hilx\
sample.  The background subtracted is also shown in this Figure.
We find no gross dependence of the luminosity function
on environment, in agreement with the results at
lower redshift from Zabludoff \& Mulchaey (\shortcite{ZM00}).
We can therefore compare the distributions of B/T in the \hilx\
and \lowlx\ clusters, including all galaxies brighter than
$\mtot=-18.2+5\log{\h50}$ ($\gtrsim 0.01 L^\ast$).  Since the redshift
distributions of the clusters (Table~1) and the luminosity functions of their constituent
galaxies are very similar, any differences seen
in these distributions must therefore be related 
(though not necessarily directly, as we discuss in \S\ref{sec-mdr}) 
to the difference in X-ray
luminosity, cluster mass or local galaxy density.

To correct for the fore- and background contamination of the clusters,
we first consider the distribution of B/T in the field galaxy sample
from the eight MDS fields, shown in the top left panel of
Figure~3.  The distribution has a strong peak at B/T$=0$, and
65$\pm$9 per cent (error from the field-to-field standard deviation)
of all galaxies have B/T\,$<0.4$; i.e., they are disk-dominated.  In the
remaining panels on the left side of Figure~3 we show the distribution
of B/T in each of the eight \lowlx\ clusters (dotted histograms).  
We correct for foreground and background
contamination by subtracting the mean field B/T distribution; the
corrected distribution is shown as the solid histograms.  In most
cases, the effect of background contamination is small, and generally
serves to reduce the fraction of disk-dominated galaxies by only a few
per cent.  In most clusters the field-corrected B/T distribution is
reminiscent of that seen for the field population, with a
comparatively flat distribution except for a pronounced peak in
several clusters at B/T\,$\sim 0$. The B/T distributions for the
individual \lowlx\ clusters are statistically consistent with the mean
distribution,  so there is no significant variation from cluster to
cluster.

\setcounter{figure}{2}
%
% Figure 2
%
\centerline{\psfig{file=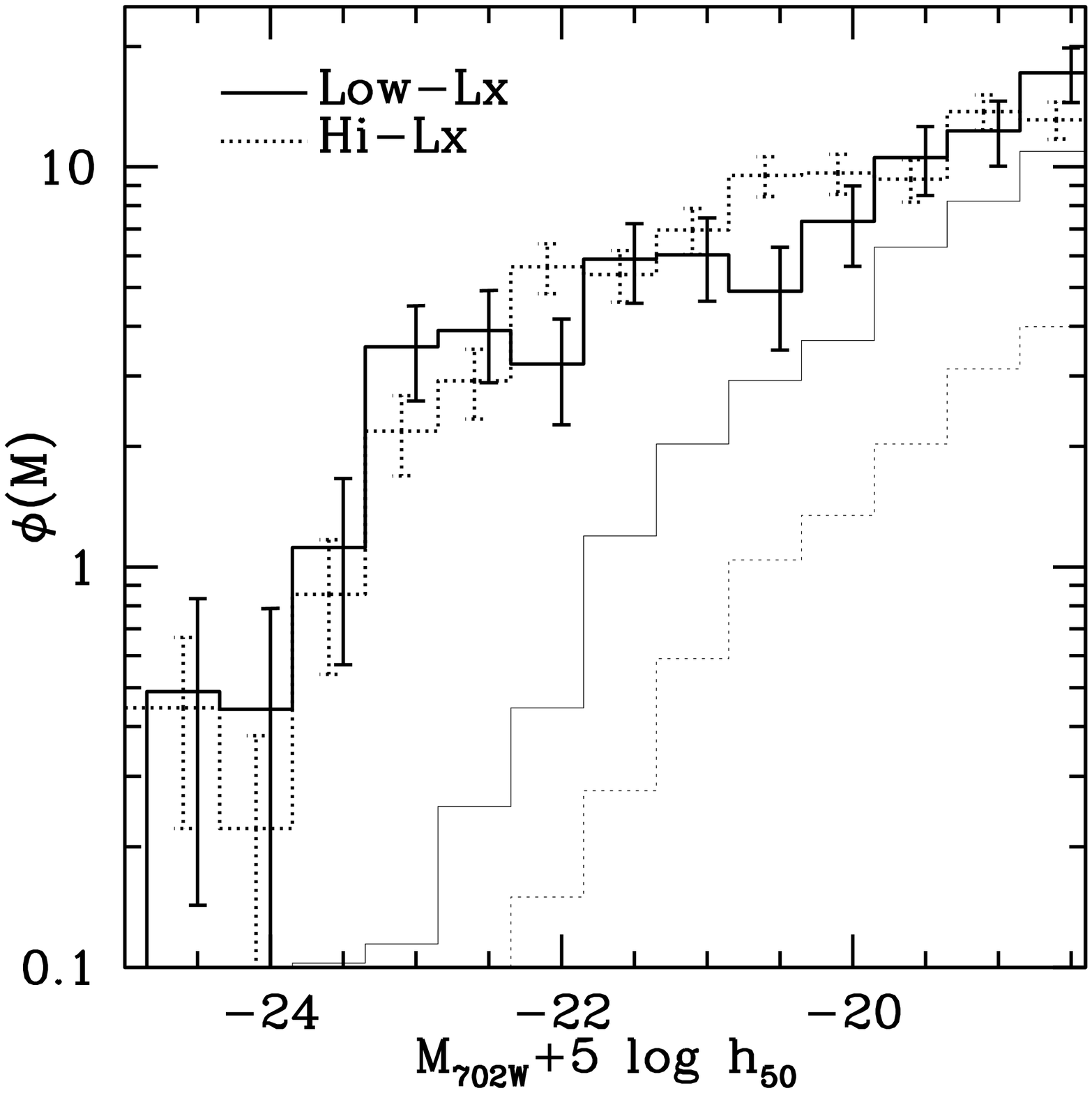,width=3.0in}}\nopagebreak
\smallskip
\noindent{\scriptsize \addtolength{\baselineskip}{-3pt} {\sc Fig. 2.}---
The luminosity functions of galaxies in the \lowlx\ {\it (heavy, solid
line)} and \hilx\ {\it (heavy, dotted line)} clusters after field
correction, both renormalized to the average number of galaxies per
cluster in the \hilx\ sample.  The error bars include the uncertainty
in the background correction; the field component subtracted from the
\lowlx\ and \hilx\ samples are shown as the {\it thin solid} and {\it
thin dotted} lines, respectively.  

}
\bigskip

%
% Figure 3
%
\begin{figure*}
\centerline{\psfig{file=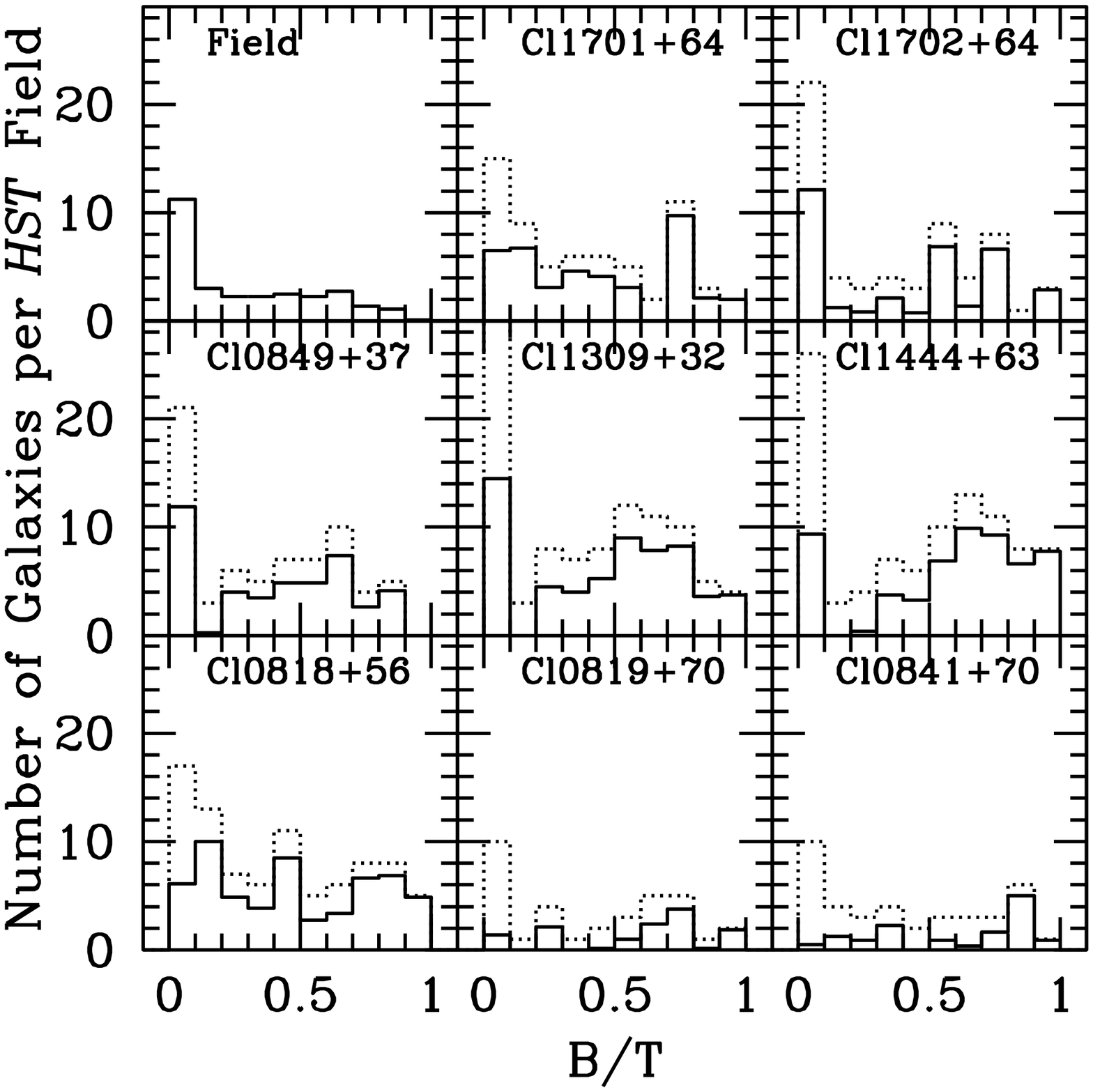,width=3.5in}\hspace*{0.5cm}\psfig{file=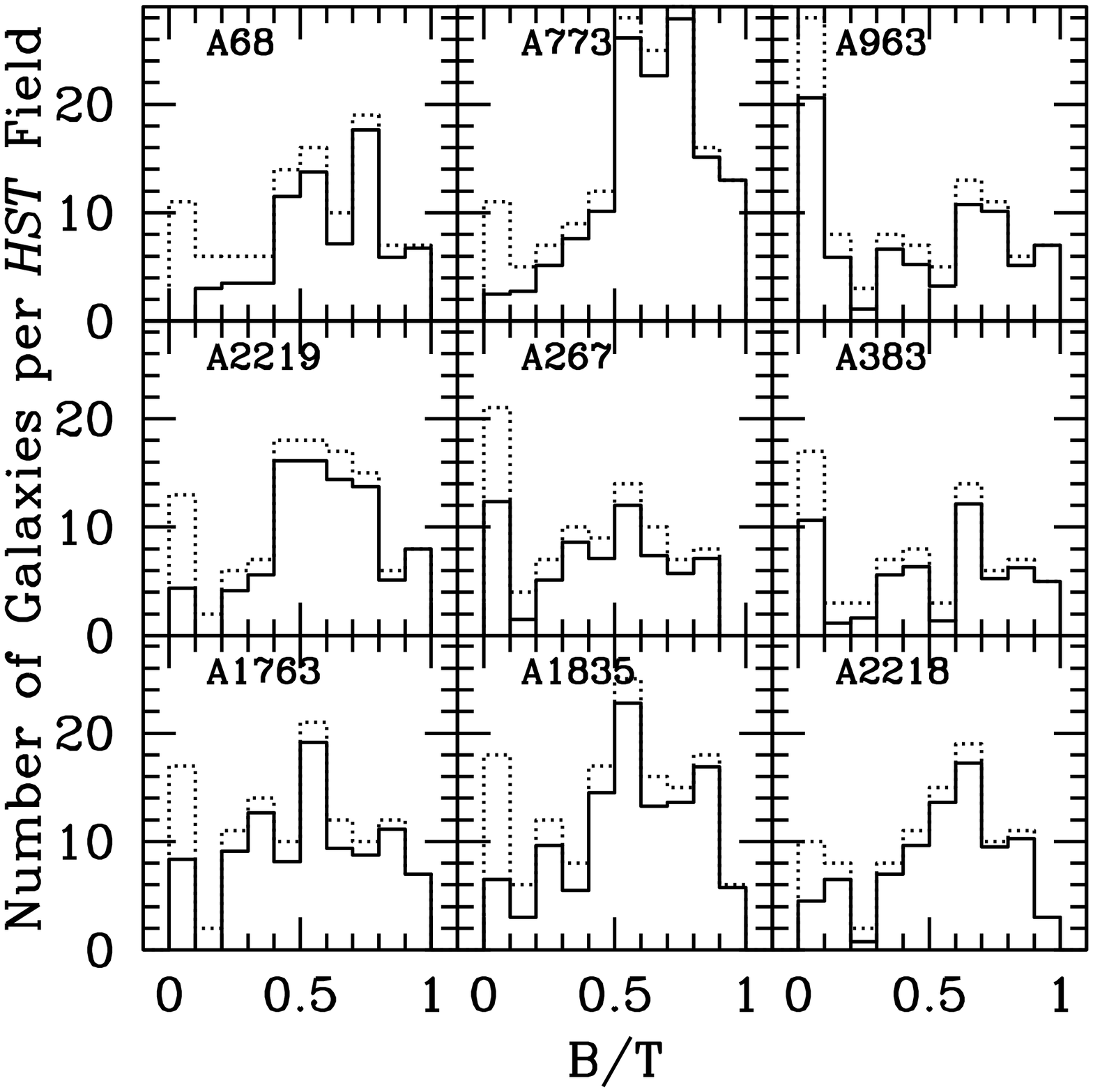,width=3.5in}}
\caption{\scriptsize \addtolength{\baselineskip}{-3pt}
{\bf Left:} The distribution of B/T in the eight
\lowlx\ clusters, and the mean distribution of the MDS field sample.
For the eight clusters, the {\it dotted} histograms show the
distribution of all galaxies, and the {\it solid} histograms are
corrected for fore- and background contamination by subtracting the
field distribution.  {\bf Right: }The B/T distributions of the \hilx\
clusters.
}
\end{figure*}

The B/T distributions for the clusters of the \hilx\ sample 
are shown in the right hand
panels of Figure~3, again with and without field correction (this is less
important for this sample than for the \lowlx\ clusters).  In general,
the \hilx\ sample shows more marked differences from the field population
than the \lowlx\ systems.  In particular, most of the clusters lack large
numbers of galaxies with B/T\,$\sim 0$ and instead show a  broad peak
of bulge-strong galaxies with B/T\,$\sim 0.5$. There are three notable
exceptions to this general statement: A\,963, A\,267 and A\,383, which all
show a peak at B/T=0, and distributions which are broadly similar to the
\lowlx\ clusters.  
%Of these only A\,963 and A\,383 have B/T distributions
%significantly different from the mean distribution of the \hilx\ clusters; 
%the difference has a
%reduced $\chi^2$ statistic of 2.3 and 1.7, respectively.  
We discuss these
three discrepant clusters in more detail in \S\ref{sec-discuss}.

The simplest way to characterise the morphological mix of the clusters
is to evaluate the fraction of disk-dominated galaxies in each cluster
after field-correction.  We arbitrarily define a disk-dominated galaxy
to have B/T$<0.4$; however, our qualitative conclusions are unchanged
for any definition ranging from B/T$<0.1$ to B/T$<0.5$.  In Figure~4
we show the fractions of disk-dominated galaxies in the clusters as
a function of their X-ray luminosities.  The disk-dominated galaxy fraction in the
field sample has a mean of 0.65 and we show this for reference; along
with the 1-$\sigma$ field-to-field standard deviation of 0.09. 
The error-weighted mean  fraction of disk-dominated
galaxies in the \hilx\ clusters is 0.23$\pm 0.10$, where the quoted uncertainty
represents the 1-$\sigma$ standard deviation from cluster to cluster.  This 10 per cent
variation is well above the statistical uncertainty, and is due to the
three clusters noted earlier (A\,963, A\,267 and A\,383) which have
unusually high fractions of disk-dominated galaxies.  For the \lowlx\
cluster sample the spread in B/T is similar, but the disk-dominated
fraction is much higher on average, at 0.40$\pm0.09$.  However, even in
these systems the disk fraction is still significantly lower than 
the general field.  This suggests that the processes influencing  galaxy
morphology can effect galaxies inhabiting structures less massive than
X-ray luminous clusters, in agreement with the results of Tran et al.\
(\shortcite{Tran}).
However, we caution that the field galaxy sample is magnitude
limited, and therefore has absolute magnitude and redshift
distributions different from those of the volume-limited cluster samples.  In particular,
we expect the redshift distribution of the field sample to peak
at $z\approx 0.55$ (\cite{CFRS5}).

%
% Figure 4
%
\centerline{\psfig{file=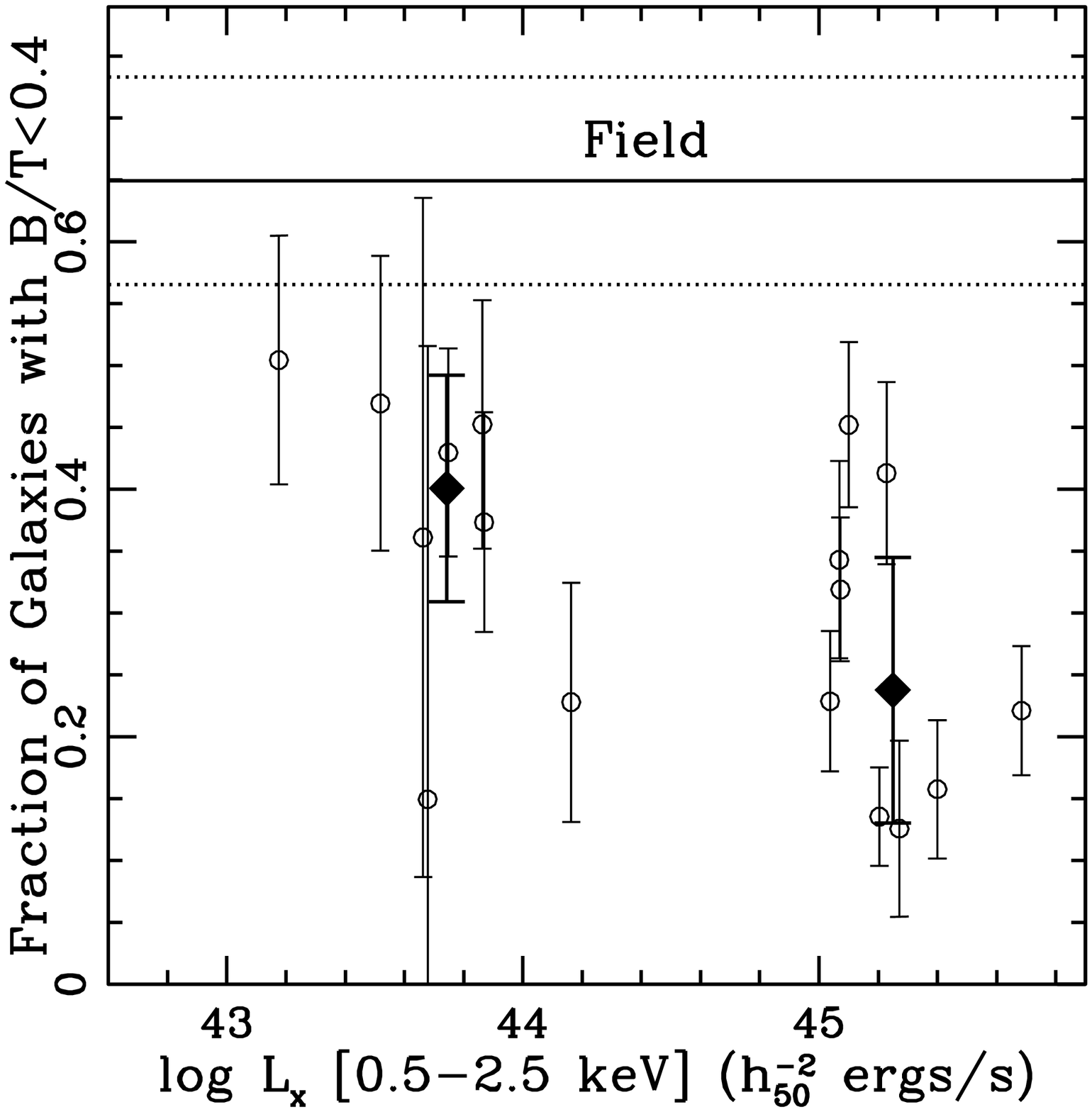,width=3.0in}}
\smallskip
\noindent{\scriptsize \addtolength{\baselineskip}{-3pt}
{\sc Fig. 4.}--- The fraction of disk-dominated galaxies (those with
B/T$<$0.4) in the \lowlx\  and \hilx\ samples as a function of the cluster
X-ray luminosities ({\it open circles}).   Error bars on the data points
are 1-$\sigma$ bootstrap estimates, and include the uncertainty in the
background correction.  The {\it solid diamonds} show the error-weighted
mean fractions for the combined \lowlx\ and \hilx\ samples, with an
error bar which represents the 1$-\sigma$ standard deviation in this
fraction from cluster to cluster (i.e.\ it is not the error on the mean).
The solid line is the corresponding fraction in the field population,
and the dashed lines show the 1-$\sigma$ field-to-field standard deviation of
this quantity.

}
\bigskip
\bigskip

\subsection{Local or global environment?}\label{sec-mdr}

The results of the previous section show that the distribution of
galaxy morphologies is different in the \lowlx\ and \hilx\ cluster
samples.  We now wish to test whether this is due to differences in
the local galaxy environment, or whether it instead reflects the influence of
the global environment on the galaxy populations.  

The clusters in the two samples we analyse have significant differences
in their physical properties, with the mean $L_X$ of the \hilx\ sample
a factor of $\sim 30$ times higher than that of the \lowlx\ sample. This
corresponds to a difference of a factor $\sim 5$ in mass, or a factor of $\sim
2$ in virial radius (from $M\propto L^{1/2}\propto R_{\rm vir}^3$; e.g.\
see Babul \etal\ \shortcite{Babul2}).
As the galaxy samples in the \hilx\ and \lowlx\
clusters are selected from the same {\it physical} region ($\sim 3\arcmin$
in diameter or   $\sim 1 \h50^{-1}$\,Mpc for a $\Lambda$CDM cosmology),
the galaxies in the \lowlx\ sample are selected from a region covering
a substantially larger fraction of the virial radius of the clusters,
compared with the \hilx\ sample.  Within relaxed systems local density
is strongly correlated with radius normalized to the virial radius
(\cite{CYE});
we therefore expect that the typical galaxy density
in the regions of the \hilx\ and \lowlx\ clusters analysed here will differ.
Thus, if a strong morphology--density relation exists in the clusters
(\cite{Dressler,PG84,Tran}),
the systematic
variation in the local galaxy density between the two samples will lead
to differences in the apparent morphological mix.

%
% Figure 5
%
\centerline{\psfig{file=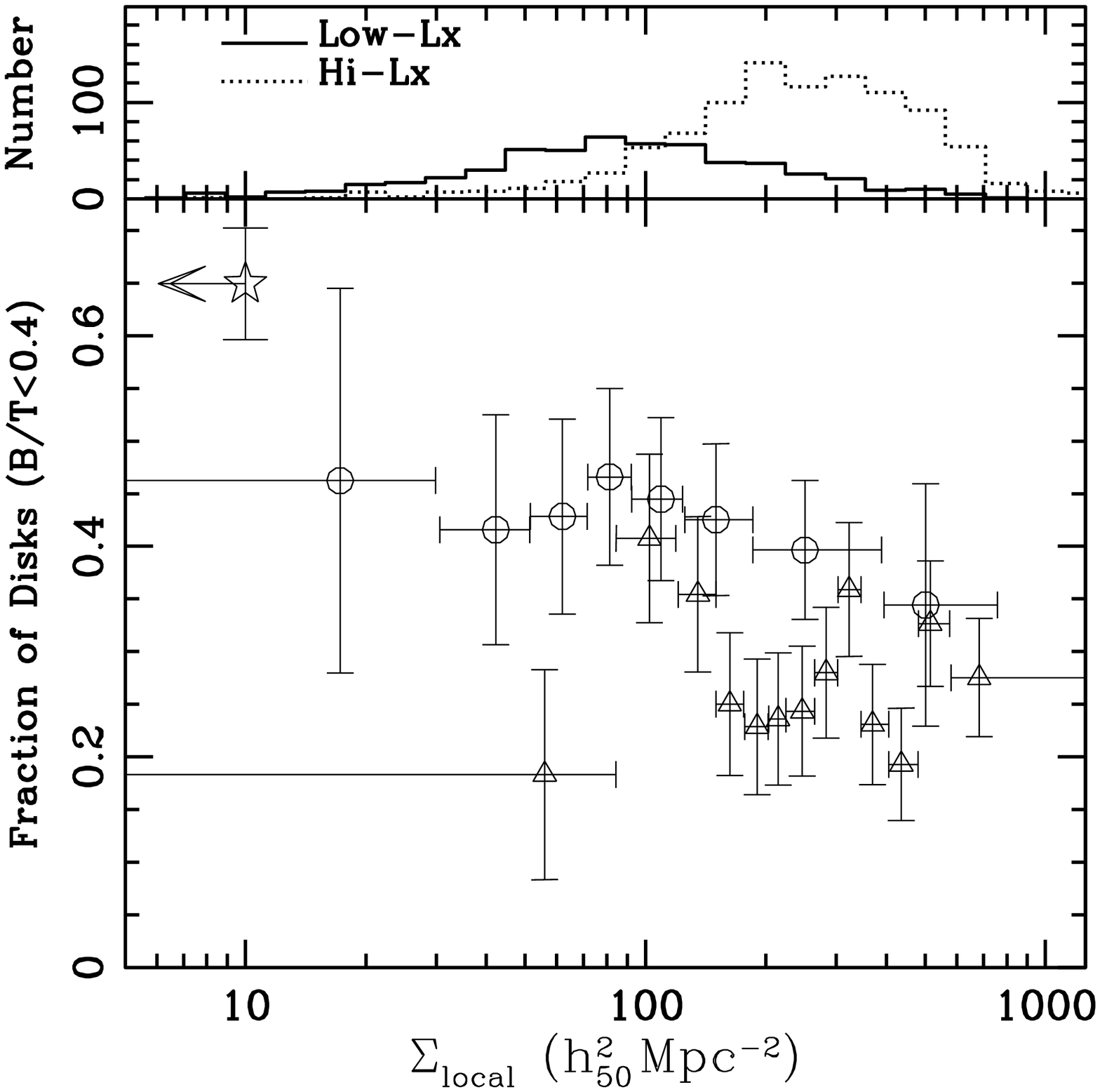,width=3.0in}}
\smallskip
\noindent{\scriptsize \addtolength{\baselineskip}{-3pt}
{\sc Fig. 5.}--- {\it Top panel: } The distribution of projected local
galaxy surface density in the \lowlx\ {\it (solid line)} and \hilx\ {\it
(dotted line)} samples.  {\it Bottom panel: } The fraction of galaxies
with B/T$<$0.4 as a function of projected local galaxy density.  The values
are corrected for background contamination as described in the text.
The field fraction (which corresponds to $\Sigma_{\rm local}\sim 0$) is
shown as the {\it star} at an arbitrary density for display purposes.
The {\it circles} and {\it triangles} correspond to the \lowlx\ and
\hilx\ samples, respectively.  Bin sizes are adaptively varied so that
each includes 75 galaxies; the horizontal error bar shows the bin span,
and the point represents the median.  Vertical error bars are determined
from bootstrap resampling which accounts for the uncertainty in the
background correction.

}
\bigskip

The local projected galaxy density around each galaxy is computed
in a manner similar to that of Dressler (\shortcite{Dressler}).
The five nearest neighbor
galaxies are found, down to the appropriate limiting magnitude (see
\S2.4) and the encompassing area is taken to be a circle which extends
out to the most distant neighbor (correctly accounting for the smaller
area available for galaxies near the boundary of the {\it WFC} mosaic).
We choose the fifth nearest neighbor, rather than the more traditional
choice of the tenth, since the latter choice results in an uncomfortably large
smoothing scale, relative to the small area of the {\it WFC} images.
The average field density  computed at the appropriate magnitude limit
($\Sigma_f=5.4$ arcmin$^{-2}$ at $I_{814}=22.5$) is subtracted to give
the local density $\Sigma_{\rm local}$, which is converted to physical
units ($\h50^{2}$\,Mpc$^{-2}$) assuming a $\Lambda$CDM cosmology.
When computing the projected densities, all of the surrounding galaxies in
the photometric catalog above the magnitude limit are considered, though
we have only determined morphologies for galaxies which are well clear of
chip boundaries.  We have not accounted for the fact that the trimmed chip
images do not exactly join, so local densities near the common boundary
of two {\it WFC} chips will be slightly underestimated.  

The contribution of the background to the observed B/T
distribution depends on the local projected density of galaxies under
consideration; in denser regions, the background correction will be
proportionally smaller.  For the total cluster sample, we computed the mean projected
field density from the MDS, $\Sigma_f$, to the appropriate magnitude
limit for each cluster, and subtracted the corresponding number of galaxies.   
For galaxies within a restricted range of
projected densities, we
scale the MDS B/T distribution appropriately so that a fraction
$f=\Sigma_f/(\Sigma_f+\bar{\Sigma})$ of the galaxy population is
statistically subtracted from the observed B/T distribution, where
$\bar{\Sigma}$ is the median, background-corrected density of the
galaxy subsample.  For
$\bar{\Sigma} \sim 25 \h50^2$ Mpc$^{-2}$, $f \sim60$ per cent,
while for $\bar{\Sigma} \sim 250 \h50^2$ Mpc$^{-2}$, $f \sim10$
per cent.  

In the top panel of Figure~5 we show the $\Sigma_{\rm local}$
distributions of the \lowlx\ and \hilx\ samples.  As expected, most of
the galaxies in the \lowlx\ sample are drawn from regions of much lower
galaxy density than seen for the bulk of the galaxies in the \hilx\ sample.
Note that the typical densities in the \hilx\ sample are much larger
than those shown in Dressler (\shortcite{Dressler}),
since we are including galaxies
about 2.5 magnitudes less luminous than in Dressler's study.

%
% Figure 6
%
\centerline{\psfig{file=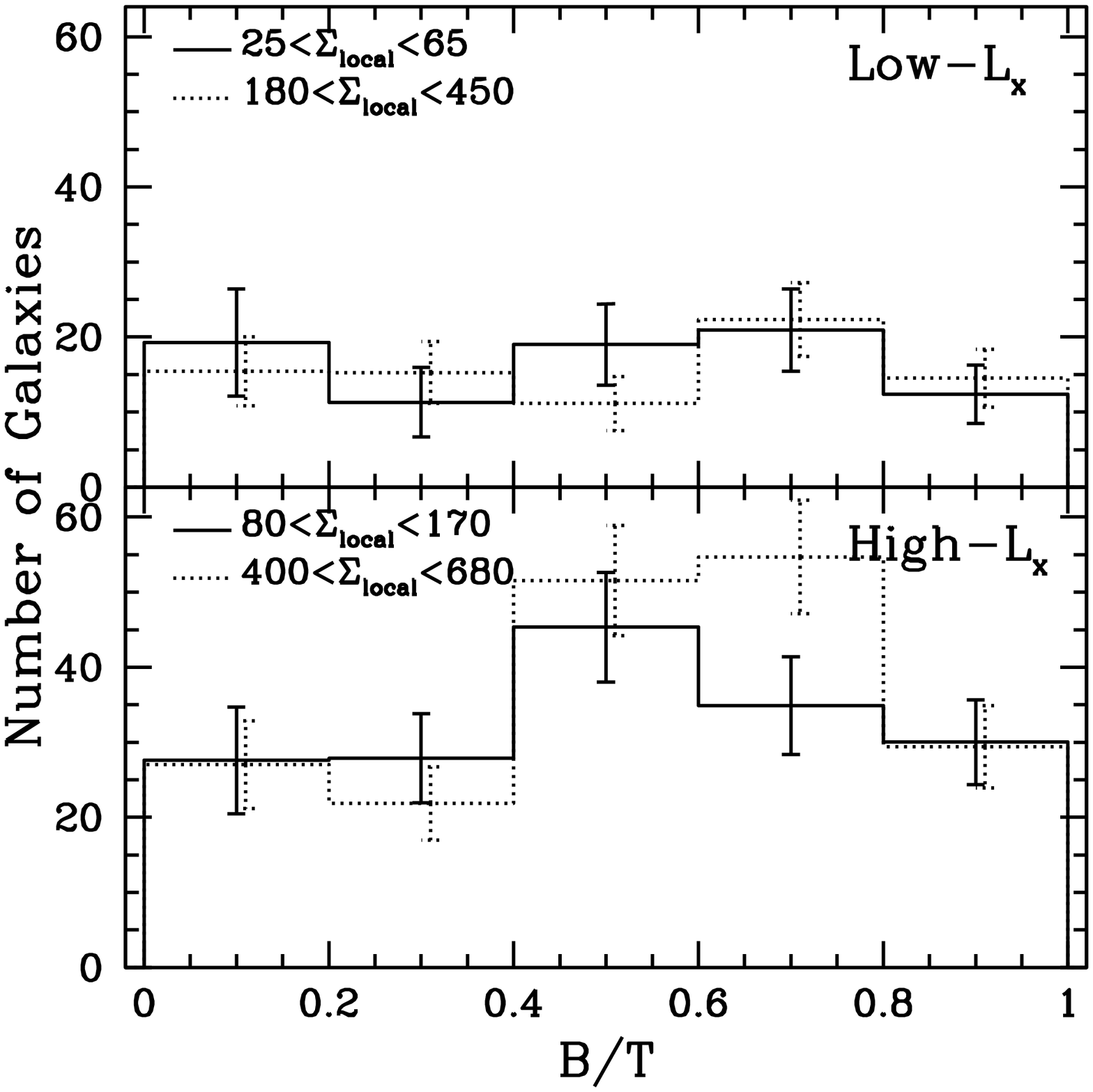,width=3.0in}}
\smallskip
\noindent{\scriptsize \addtolength{\baselineskip}{-3pt}
{\sc Fig. 6.}--- The distribution of B/T ratios is shown for low (lowest
5-25 percentile) and
high (highest 5-25 percentile) 
density regions separately, as indicated.  The \lowlx\ cluster data are shown in
the {\it top panel}, and the \hilx\ cluster data are shown in the {\it
bottom panel}.   Uncertainties are 1$-\sigma$ and include the error in
background subtraction.

}
\bigskip

The morphology-density relation is shown in the bottom panel of Figure~5.
We use an adaptive bin size so that each bin contains 75 galaxies, and
compute the field-corrected fraction of those that are disk-dominated,
B/T$<0.4$ (as this definition of disk-dominated is arbitrary we will
consider the full B/T distribution as a function of density below).
In the individual cluster samples there is only weak evidence for
a morphology--density relation due to the small samples and limited
dynamic range in density.  In Figure~6 we compare the background-subtracted distribution
of B/T in low (5-25 percentile) and high (75-95 percentile) density regions within the individual \lowlx\
and \hilx\ samples.  
The corresponding density limits in each sample are also shown in Figure~6.  This
compares galaxies in well separated density ranges above and below the mean of the sample,
while avoiding
the extreme tails of the distribution.  
%For the \lowlx\ clusters, each subsample
%contains about 80 galaxies, while for the \hilx\ clusters there are about
%160 galaxies in each distribution.  
No statistically significant difference
can be seen between the low and high density regimes, in either cluster
sample.  There is an indication, in the \hilx\ clusters, that 
the lowest density galaxies are distributed toward lower B/T, but
the reduced $\chi^2$ statistic for the difference between
these two distributions, accounting for the uncertainty in background
subtraction, is only 1.2\footnote{We are forced to bin the data in order to
perform the background subtraction, and therefore we use the $\chi^2$ 
statistic, rather than the Kolmogorov-Smirnov statistic.  The disadvantage
is that this statistic is quite sensitive to the chosen binning, and we
do not, therefore, consider this a very robust estimate of the
significance of our result.  This is true of all estimated likelihoods
based on the $\chi^2$ statistic, throughout this paper.}.  The \MDR\ is clearly weak,
at best; this can be attributed partly to the limited
dynamic range of densities explored; {\it HST} imaging over a wider field
would allow the extension of the analysis of the \hilx\ clusters to much
lower densities, and improve the comparison.

There is some evidence from Figure~5 that, at a given overdensity,
disk-dominated galaxies are more common in the \lowlx\ clusters.  We
investigate this possibility by restricting the galaxy
samples in both \hilx\ and \lowlx\ clusters to a common density range
where both samples have good statistics, $\Sigma_{\rm local}=50$--200.
In this density range, there is good overlap between the two datasets,
and limited gradient in morphological composition.  Although this
density range is still fairly large, our results do not qualitatively change
as this interval is reduced, at the necessary expense of
lower statistical significance.

The B/T distributions in the \hilx\ and \lowlx\ clusters, restricted
to $\Sigma_{\rm local}=50$--200, are shown in Figure~7.  The two distributions appear to
differ, with a relative excess of galaxies in the \hilx\ sample with
B/T$ \sim$0.5, and a corresponding deficit of those with B/T\,$\sim 0$.
A statistical comparison of the two normalised distributions gives a reduced $\chi^2$
of 3.1 and a corresponding likelihood of 98\%.  This difference is much
more pronounced and significant (99.99\% likelihood) if we exclude the three unusual \hilx\
clusters discussed in \S\ref{sec-mdist}; however, we are reluctant to do so without 
identifying the origin of the difference between these and the majority
of the \hilx\ systems.  We also reemphasize that the $\chi^2$
test is sensitive to the binning of the data and thus does not provide
a complete statistical description of the difference between the samples.

\section{Discussion}\label{sec-discuss}
From Figures 3 and 4, it is evident that all of the \lowlx\ clusters have 
similar galaxy morphology distributions; all eight clusters are dominated
by galaxies with B/T$\approx0$, and are statistically indistinguishible
from the mean B/T distribution of the combined sample.  Most of the \hilx\
clusters differ from this distribution, as they 
are dominated by galaxies with intermediate B/T.  The exceptions are three
clusters, A\,963, A\,267 and A\,383, which have B/T distributions more similar
to those found in the \lowlx\ clusters.  
There is nothing
strikingly unusual about the morphology of these three clusters; all are
dominated by a single cD galaxy and do not show obvious signs of irregularity.  The redshifts
of these three clusters are not at either extreme of the distribution, nor are the
galaxy luminosity functions, or distributions of local projected density, 
significantly different from that of the average 
\hilx\ cluster.

One possible explanation for the difference could be that these three clusters
are less massive than the other \hilx\ clusters, and that their X-ray luminosities
are unusually high, or overestimated.  Alternatively, if the clusters have an
unusually large number density of fore- and background galaxies along the line of 
sight, the background correction will have been underestimated.  
However, the local galaxy density
distribution of the three clusters is indistinguishible from the \hilx\ cluster average;
that is, the galaxy density is as high as we expect for the cluster's luminosity.
It would seem, therefore, that {\it both} the luminosity and the density would have
to be artificially enhanced for these explanations to work; this would
be an uncomfortable coincidence.  
It therefore appears
that galaxy morphology is sensitive to factors other than just local density or
X-ray luminosity, though it is not clear what they are.  

%
% Figure 7
%
\centerline{\psfig{file=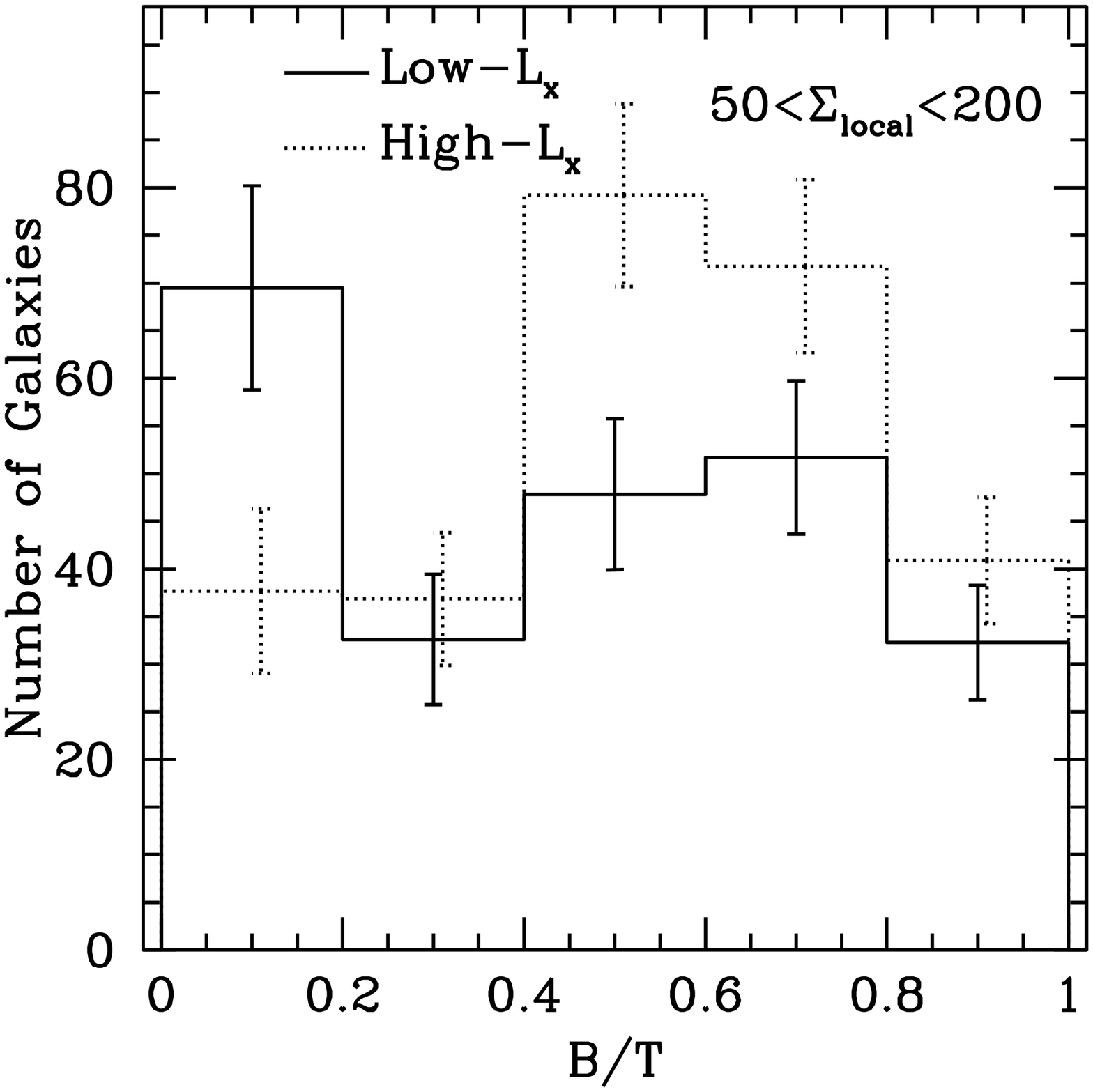,width=3.0in}}
\smallskip
\noindent{\scriptsize \addtolength{\baselineskip}{-3pt}
{\sc Fig. 7.}--- The B/T distribution for galaxies with $\Sigma_{\rm
local}=50$--200\,$\h50^{2}$\,Mpc$^{-2}$.  The {\it solid} and
{\it dotted} histograms correspond to the \lowlx\ and \hilx\ samples,
respectively.

}
\bigskip
\setcounter{figure}{7}

Putting aside for the moment the intrinsic variation in the \hilx\ 
sample, we can explore differences in the average properties of the
\hilx\ and \lowlx\ clusters.  There is some evidence that disk-dominated
galaxies are relatively more common in the \lowlx\ clusters, at fixed
overdensity, relative to the \hilx\ clusters (Figure~7).  Although this result is
of marginal statistical significance, it is still of interest to
investigate which
physical process could give rise to the effect.  In currently popular
models of galaxy formation, bulge-dominated galaxies are generally
created in one of two ways.  The first is by the merger of two
galaxies, in which some or all of the baryonic material (from the
progenitor bulges and disks) is converted into a new bulge component.
In this case, differences in the bulge luminosity function will
reflect differences in the galaxy merger histories.  Alternatively, a
bulge-dominated galaxy can be created by reducing the stellar mass of
the disk component, either by destroying it (for example, through
harassment) or by causing it to fade sufficiently following the
cessation of star formation (as might be expected if ram pressure
stripping removes all the gas from the disk).  In this case, the bulge
luminosity function should be unaffected by the efficiency of the
destruction process, but the B/T measure will increase as the disk fades
or is destroyed.  
We will attempt to distinguish between these two
possibilities by examining the relationship between bulge luminosity
and B/T in the two cluster samples.
%
% Figure 8
%
\begin{figure*}
\centerline{\psfig{file=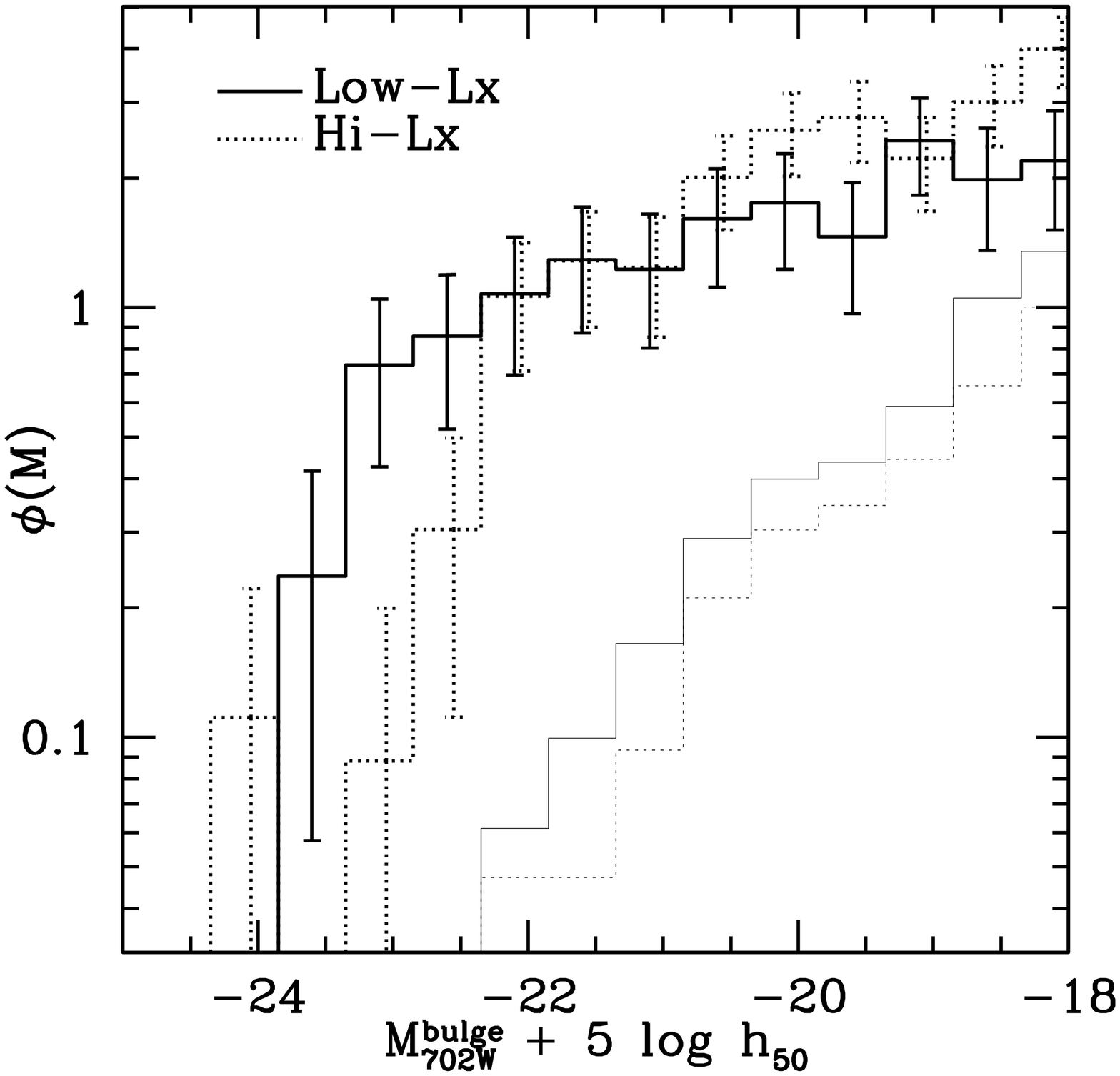,width=3.5in}\hspace*{0.5cm}\psfig{file=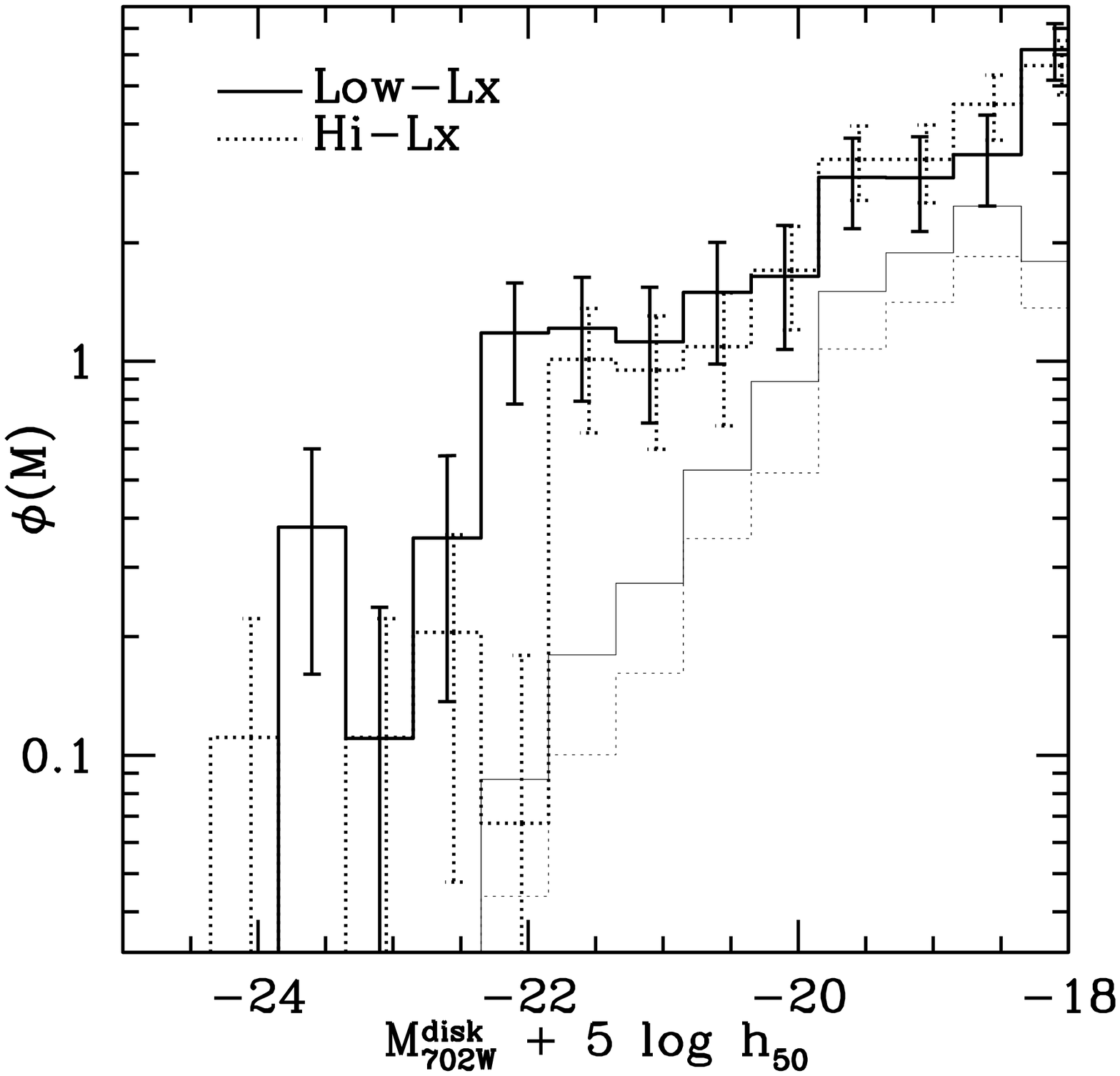,width=3.5in}}
\caption{\scriptsize \addtolength{\baselineskip}{-3pt}
{\bf Left:} The luminosity function of the bulge component, in
the \lowlx\ sample ({\it heavy, solid line}) and the \hilx\ sample
({\it heavy, dotted line}), restricted to regions of overdensity
$50<\Sigma_{\rm local}<200$.  $\phi$(M) is normalised to the average
number of galaxies per cluster in the \hilx\ sample, restricted to this
density range.  The thin {\it solid} and {\it dotted} lines show the
subtracted background component.  {\bf Right: }The same, but for the
disk components.  }
\end{figure*}

We show the bulge and disk luminosity functions separately in Figure~8,
for those galaxies from \lowlx\ and \hilx\ samples lying within the
density range $\Sigma_{\rm local}=50$--200.
The luminosity functions are normalized to the average number of galaxies
in this density range per cluster in the \hilx\ sample.  While the
disk luminosity functions appear to be very similar in both cluster
samples, there is an indication (not statistically significant, however)
that the bulge luminosity function
is steeper in the \hilx\ clusters; i.e., there is a higher proportion of
low-luminosity bulges compared with the \lowlx\ sample.  
This suggests that the difference in B/T distributions is reflecting a difference
in the properties of the bulges, rather than the disks.  With some effort, this
can be seen more clearly in the correlation of B/T with disk or bulge
luminosity, shown in Figure~9.  We again
restrict the galaxy population to those   with 
$\Sigma_{\rm local}=50$--200, to minimize morphology-density effects.
Concentrating first on the top panels, we see the correlation between
B/T and bulge luminosity, for the \lowlx\ and \hilx\ samples separately.  
Galaxies with B/T$<0.1$ have been arbitrarily
plotted at B/T$=0.1$ to make them visible in this logarithmic plot.
Note also that the shape of the sample luminosity limit, combined with
the shape of the luminosity function, gives rise to some of the structure
seen in this figure.  It is clear (as it is in Figure~7) that the \lowlx\
clusters are dominated by galaxies with low B/T, while this is much less
the case in the \hilx\ clusters.  We can now explore the two scenarios described
above.  If galaxy disks are preferentially stripped, or otherwise fade, in the \hilx\ clusters,
the galaxy density will be shifted directly upward in this figure, relative 
to the \lowlx\ distribution (see the arrows labelled ``S'').  
This would serve to push many of the galaxies with
B/T$<0.2$, which are mostly faint, below our luminosity limit.
In this respect, such a mechanism could provide a viable explanation of the
difference in galaxy distributions seen in these top figures.
However, if we now look at the bottom panels,
it can be seen that this explanation is unlikely.  In these figures, we plot {\it disk}
luminosity as the x-axis.  Here, it is clear that the faint end of the
disk luminosity function is primarily defined by galaxies with B/T$<0.2$.
If these disks were to fade beyond the luminosity limit, the disk luminosity
function would necessarily become much shallower, and this does not appear
to be the case (Figure ~8).  Instead, the faint end of the disk luminosity
function in the \hilx\ clusters is relatively more populated by galaxies with B/T$\sim0.5$
and disk luminosities roughly equivalent to the disk luminosities of the B/T$\sim0$ population
in the \lowlx\ clusters.

On the other hand, if the difference between
the \lowlx\ and \hilx\ clusters is due to the fact that the {\it bulges}
are systematically brighter in the \hilx\ clusters, due perhaps to a more
extensive merger history, galaxies in these panels will move diagonally, as indicated
by the arrows labelled ``M''.  This explanation appears consistent with the
observations.  The B/T$<0.2$ population which dominates the \lowlx\ clusters
can be translated in the sense of brightening their bulges at constant disk luminosities,
to reproduce the distribution seen in the \hilx\ clusters.

We conclude that is unlikely the B/T$\sim 0.5$ galaxies in the \hilx\ clusters have been formed
from a process which operates uniquely in this environment by 
stripping or otherwise destroying the disk in a lower B/T system.
Rather, the results seem to suggest that there are differences in
bulge growth in the \lowlx\ and \hilx\ samples.  This  provides
quantitative support for the suggestion by Dressler (\shortcite{Dressler}), 
that the early-type galaxies in clusters cannot all be formed by destroying the
disks of a normal population of late-type galaxies.
However, this is not evidence that ram pressure stripping (or other
processes with similar effect) do not take place at all; at least, they could
play a strong role in establishing the local morphology-density
relation.  But the difference between the \MDR\ for \lowlx\ and
\hilx\ clusters indicates that there are other processes at work.

%
% Figure 9
%
\begin{figure*}
\centerline{\psfig{file=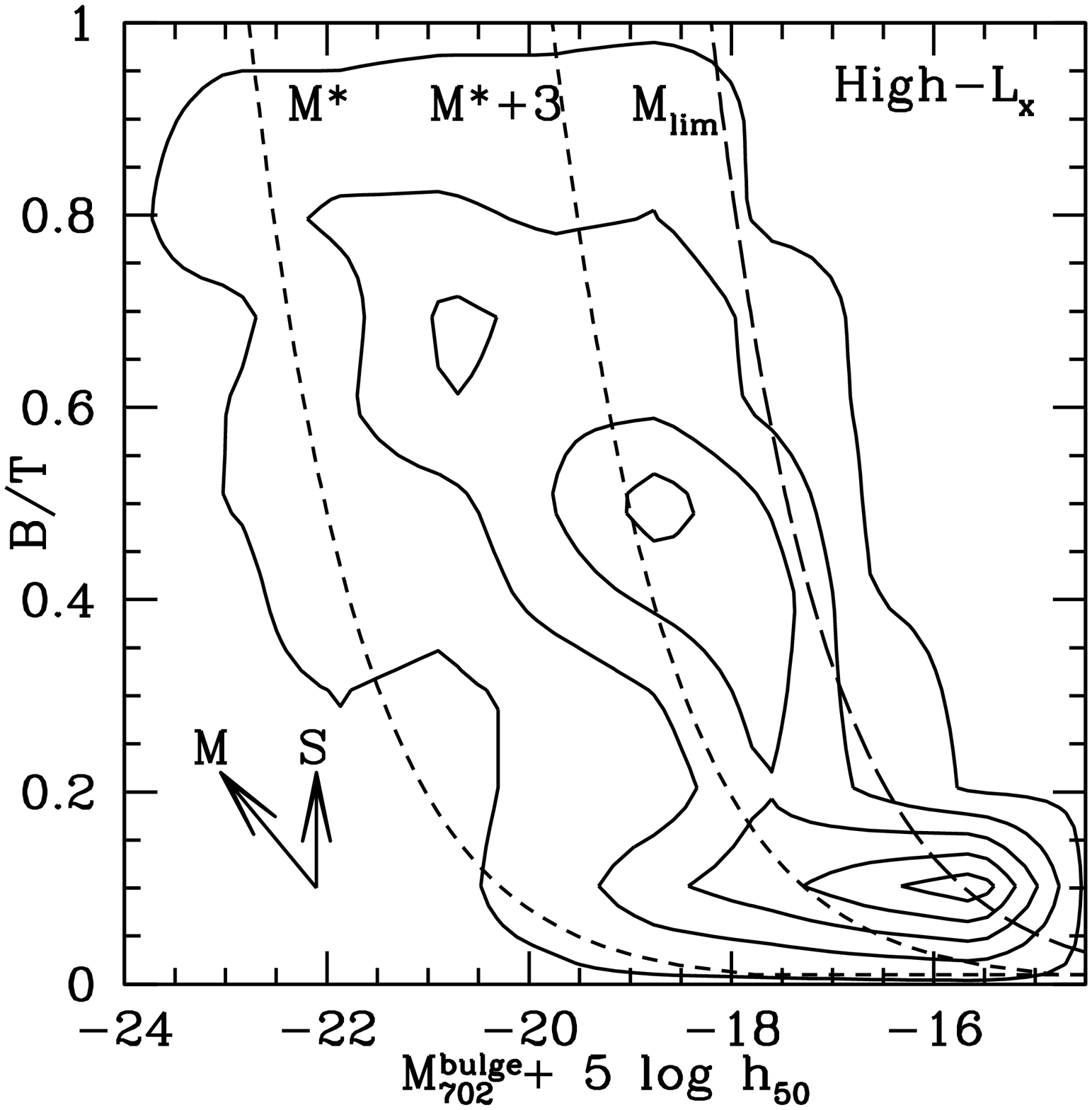,width=3.5in}\hspace*{0.5cm}\psfig{file=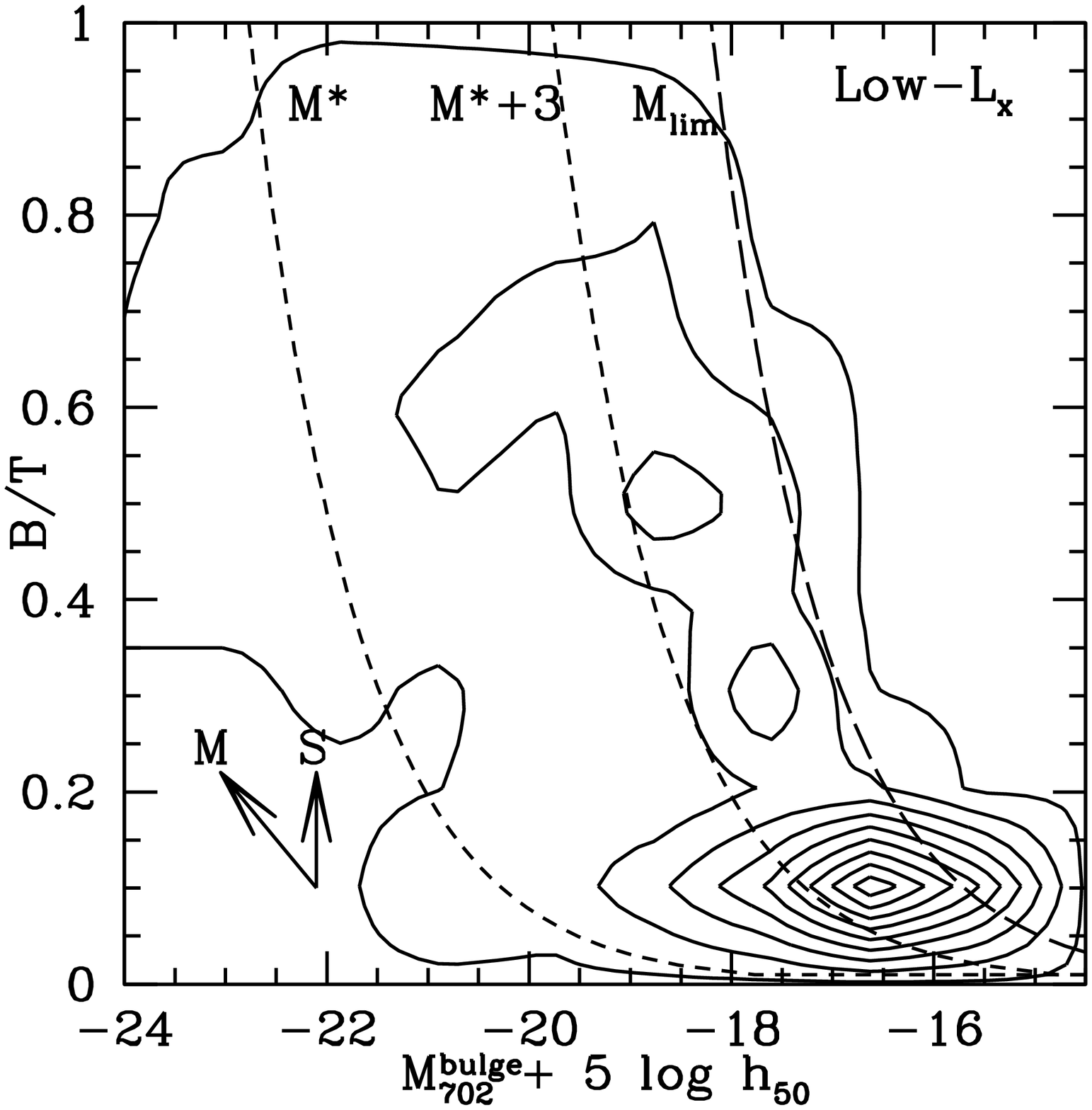,width=3.5in}}
\centerline{\psfig{file=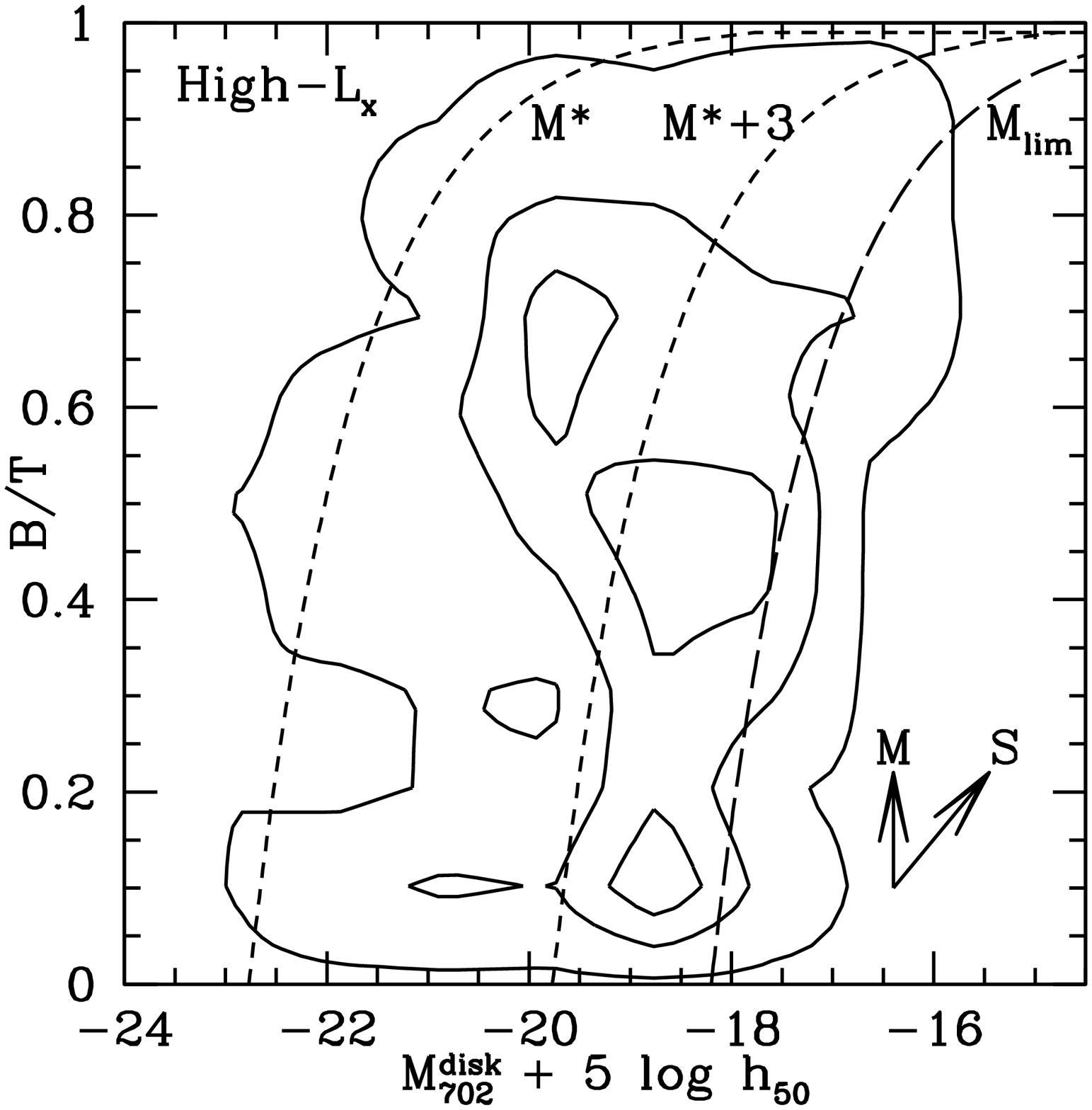,width=3.5in}\hspace*{0.5cm}\psfig{file=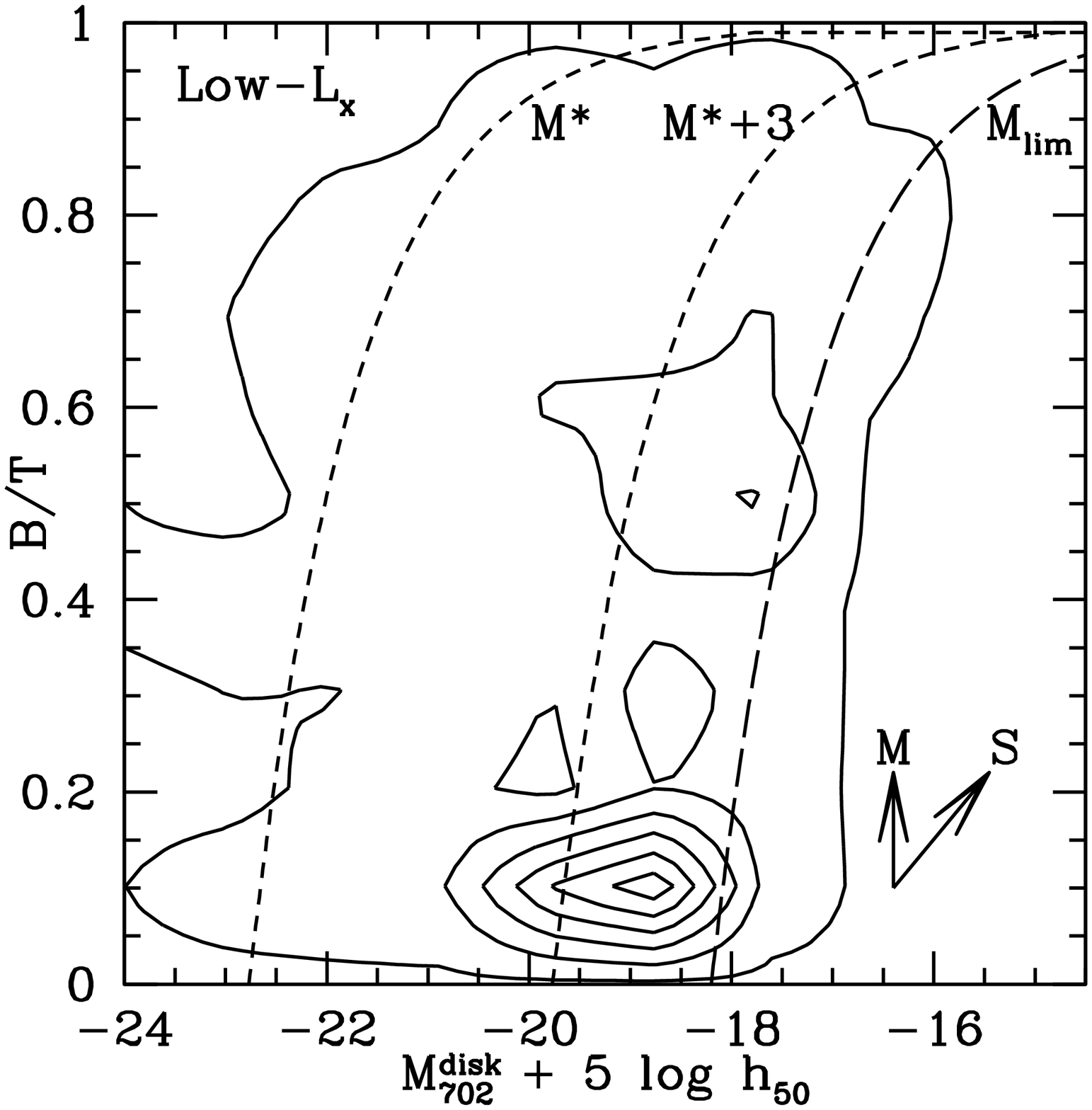,width=3.5in}}
\caption{\scriptsize \addtolength{\baselineskip}{-3pt}
{\bf Top:} The correlation between B/T and bulge luminosity is shown
for galaxies in the \hilx\ clusters {\it (left)} and the \lowlx\ clusters
{\it(right)}, after background subtraction.  The sample
is restricted to those galaxies in the local projected density range
$\Sigma_{\rm local}=50$--200\,$\h50^{2}$\,Mpc$^{-2}$.
The density of galaxies in this plane is represented as a
contour plot; the contours are equally spaced in intervals of 5
galaxies per cell (starting with 1), where a cell is 1 mag and 0.1
in B/T.  Galaxies with B/T$<0.1$ are arbitrarily fixed at B/T=0.1
to be visible in these plots.  
The {\it short-dashed} lines show the bulge luminosity
of an M$^\ast$ galaxy, and a
galaxy three magnitudes fainter, as a function of B/T, for reference.
The {\it long-dashed} line represents the luminosity limit of the sample.
The arrows labelled ``S'' and ``M'' show the magnitude and direction a galaxy with
B/T=0.1 would move, if its disk faded by one magnitude (stripping: S) or its bulge brightened
by one magnitude (merging: M).  
{\bf Bottom: }
Same as the top panel, but for the correlation between B/T and
disk luminosity.  Galaxies with B/T$>0.9$ are arbitrarily fixed at
B/T$=0.9$ to be visible in these plots.  
} \nocite{Sloan_lf}
\end{figure*}

%
% Figure 10
%
%\begin{figure*}
%\centerline{\psfig{file=Fig10.ps,width=7.in,angle=270}}
%\caption{\scriptsize \addtolength{\baselineskip}{-3pt}
%Twelve galaxies selected from the \hilx\ clusters, populating
%the peak of the distribution in the bottom left panel of Figure~9.
%Total luminosities and B/T values are shown in each panel.  Each image
%is 6\arcsec\ on a side, and contours are
%arbitrarily spaced.
%These galaxies are a subsample of the population which appears to be
%in excess in the \hilx\ clusters, relative to the \lowlx\ clusters.
%}
%\end{figure*}

\section{Conclusions}\label{sec-conc}

We have presented an {\it HST}-based  morphological analysis of
galaxies in clusters from two samples at $z \sim0.25$.  The two data
sets are extremely well matched in the properties and quality of their
data; all observations are 3-orbit {\it HST} exposures in the F702W
filter, and both cluster samples have a similar mean redshift and
small redshift range, thus minimizing differences in k-corrections and
absolute magnitude limits.  The only significant difference between
the samples is the X-ray luminosity of the clusters contained in them; 
the clusters in the \lowlx\
sample have $L_X$(0.1--2.4\,keV)=0.15--1.5$\times
10^{44}$\,ergs\,s$^{-1}$, while the \hilx\ clusters have
$L_X$(0.1--2.4\,keV)$\geq 10^{45}$\,ergs\,s$^{-1}$.   These data are
therefore ideal for testing the dependence of galaxy morphology on
local density and the global environment.  We summarize our findings
as follows:

\begin{itemize}

\item Within a fixed physical region ($\sim 1 \h50^{-1}$ Mpc diameter), the
fraction of disk-dominated galaxies is strongly dependent on $L_X$.  In
the \hilx\ sample, the fraction of galaxies with B/T$<0.4$ is 0.25$\pm
0.10$, while in the \lowlx\ sample it is 0.44$\pm 0.07$.

\item In both cluster samples, the fraction of galaxies with B/T$<0.4$
is significantly lower than the fraction of such galaxies in the field,
0.65$\pm0.09$ (where the uncertainty represents the field-to-field
variation).  The origin of this discrepancy is, however, unclear
because the comparison is between a volume-limited sample at
$z\sim0.25$ (the
clusters) and a magnitude-limited one with a broad redshift
distribution peaked at $z\sim 0.55$ (the field).

\item There is a dependence of the typical B/T of a galaxy on its local
projected galaxy density. This is partly responsible for the above
result, as galaxies in the \hilx\ clusters are generally located in
denser environments.  However, when we compare galaxies at similar local
densities in the \lowlx\ and \hilx\ samples, small differences persist,
with the \hilx\ clusters showing an excess of bulge-strong galaxies compared
to the \lowlx\ sample.  This difference is moderately significant, with
confidence limits of 98\% (as determined by a $\chi^2$ test, which is sensitive to how
the data are binned).

\item To investigate this behaviour in more detail we focus on the bulge
luminosity function in the two samples.  We find some evidence for a
difference in the bulge luminosity function of galaxies between the
\lowlx\ and \hilx\ clusters when we restrict the comparison to galaxies
with local densities between 50 and 200 $\h50^{2}$\,Mpc$^{-2}$,
although the difference is not highly statistically significant.  

\item Three of the \hilx\ clusters show B/T distributions which are more similar
to the average \lowlx\ B/T distribution than the
majority of \hilx\ clusters.  Galaxy morphology therefore appears to depend on
another, unknown parameter, besides local density and X-ray luminosity.
\end{itemize}  

These observations suggest that the factors influencing galaxy
morphlogy are: 1) efficient in relatively low-mass clusters; and 2)
influence the luminosity of the bulge component, and not the disk
alone.  The most likely interpretation in our opinion is that the
merger histories of galaxies in low and high mass clusters are
different, in that merging has played a larger role for galaxies 1--3
magnitudes below $\sim L^\ast$ in more massive clusters.  This might
be expected from models of hierarchical cluster growth (\cite{LC94,K96}),
since the merger history of massive clusters
tends to be more extended.  On the other hand, galaxy-galaxy mergers
are uncommon in regions of high velocity dispersion, and it is
unclear what variation in the bulge luminosity
function would be predicted by the models.

\section*{Acknowledgements}

We thank Julio Navarro, Guinevere Kauffmann and Luc Simard for
helpful discussions about the implications of this work.  We
also appreciate comments from Bianca Poggianti, Stefano Andreon,
and an anonymous referee, which helped us improve this paper.
Financial
support is acknowledged from PPARC (MLB, RGB, GPS, RLD), the Royal
Society (IRS) and Leverhulme Trust (IRS, RLD), the Deutsche
Forschungsgemeinschaft and the VW foundation (BLZ) and CNRS (JPK).  We
also acknowledge support from the UK-French ALLIANCE collaboration
programme no.  00161XM and STScI grant GO-08249.

\bibliographystyle{astron_mlb}
\bibliography{ms}

\begin{thebibliography}{}

\bibitem[\protect\astroncite{Abell}{1965}]{A65}
Abell, G.~O.: 1965,
\newblock {\em ARA\&A} {\bf 3}, 1

\bibitem[\protect\astroncite{{Abraham} et~al.}{1996a}]{A+96}
{Abraham}, R.~G., {Tanvir}, N.~R., {Santiago}, B.~X., {Ellis}, R.~S.,
  {Glazebrook}, K., and {van den Bergh}, S.: 1996a,
\newblock {\em MNRAS} {\bf 279}, L47

\bibitem[\protect\astroncite{{Abraham} et~al.}{1996b}]{MDS-morph2}
{Abraham}, R.~G., {van den Bergh}, S., {Glazebrook}, K., {Ellis}, R.~S.,
  {Santiago}, B.~X., {Surma}, P., and {Griffiths}, R.~E.: 1996b,
\newblock {\em ApJS} {\bf 107}, 1

\bibitem[\protect\astroncite{{Allen} and {Fabian}}{1998}]{AF}
{Allen}, S.~W. and {Fabian}, A.~C.: 1998,
\newblock {\em MNRAS} {\bf 297}, L57

\bibitem[\protect\astroncite{{Andredakis}}{1998}]{A98}
{Andredakis}, Y.~C.: 1998,
\newblock {\em MNRAS} {\bf 295}, 725

\bibitem[\protect\astroncite{{Andreon}}{1998}]{Andreon98}
{Andreon}, S.: 1998,
\newblock {\em ApJ} {\bf 501}, 533

\bibitem[\protect\astroncite{Babul et~al.}{2001}]{Babul2}
Babul, A., Balogh, L., M., Lewis, G.~F., and Poole, G.~B.: 2001,
\newblock {\em MNRAS} pp in press, astro--ph/0109329

\bibitem[\protect\astroncite{Bahcall}{1977}]{B77}
Bahcall, N.~A.: 1977,
\newblock {\em ApJL} {\bf 218}, 93

\bibitem[\protect\astroncite{Balogh et~al.}{1999}]{PSG}
Balogh, M.~L., Morris, S.~L., Yee, H. K.~C., Carlberg, R.~G., and Ellingson,
  E.: 1999,
\newblock {\em ApJ} {\bf 527}, 54

\bibitem[\protect\astroncite{{Balogh} et~al.}{2000}]{infall}
{Balogh}, M.~L., {Navarro}, J.~F., and {Morris}, S.~L.: 2000,
\newblock {\em ApJ} {\bf 540}, 113

\bibitem[\protect\astroncite{Balogh et~al.}{1998}]{B+98}
Balogh, M.~L., Schade, D., Morris, S.~L., Yee, H. K.~C., Carlberg, R.~G., and
  Ellingson, E.: 1998,
\newblock {\em ApJL} {\bf 504}, 75

\bibitem[\protect\astroncite{{Bardeen} et~al.}{1986}]{BBKS}
{Bardeen}, J.~M., {Bond}, J.~R., {Kaiser}, N., and {Szalay}, A.~S.: 1986,
\newblock {\em ApJ} {\bf 304}, 15

\bibitem[\protect\astroncite{{Barnes}}{1992}]{Barnes}
{Barnes}, J.~E.: 1992,
\newblock {\em ApJ} {\bf 393}, 484

\bibitem[\protect\astroncite{Baugh et~al.}{1996}]{semianal}
Baugh, C.~M., Cole, S., and Frenk, C.~S.: 1996,
\newblock {\em MNRAS} {\bf 283}, 1361

\bibitem[\protect\astroncite{Bertin and Arnouts}{1996}]{sextractor}
Bertin, E. and Arnouts, S.: 1996,
\newblock {\em A\&AS} {\bf 117}, 393

\bibitem[\protect\astroncite{{Blanton} et~al.}{2001}]{Sloan_lf}
{Blanton}, M.~R., {Dalcanton}, J., {Eisenstein}, D., {Loveday}, J., {Strauss},
  M.~A., {SubbaRao}, M., {Weinberg}, D.~H., and {the Sloan collaboration}:
  2001,
\newblock {\em AJ} {\bf 121}, 2358

\bibitem[\protect\astroncite{{Bravo-Alfaro} et~al.}{2000}]{Bravo+00}
{Bravo-Alfaro}, H., {Cayatte}, V., {van Gorkom}, J.~H., and {Balkowski}, C.:
  2000,
\newblock {\em AJ} {\bf 119}, 580

\bibitem[\protect\astroncite{{Brinchmann} et~al.}{1998}]{Brinch98}
{Brinchmann}, J., {Abraham}, R., {Schade}, D., {Tresse}, L., {Ellis}, R.~S.,
  {Lilly}, S., {Le Fevre}, O., {Glazebrook}, K., {Hammer}, F., {Colless}, M.,
  {Crampton}, D., and {Broadhurst}, T.: 1998,
\newblock {\em ApJ} {\bf 499}, 112

\bibitem[\protect\astroncite{{Carlberg} et~al.}{1997}]{CYE}
{Carlberg}, R.~G., {Yee}, H. K.~C., and {Ellingson}, E.: 1997,
\newblock {\em ApJ} {\bf 478}, 462

\bibitem[\protect\astroncite{{Cole}}{1991}]{Cole91}
{Cole}, S.: 1991,
\newblock {\em ApJ} {\bf 367}, 45

\bibitem[\protect\astroncite{{Cooray}}{1999}]{Cooray}
{Cooray}, A.~R.: 1999,
\newblock {\em MNRAS} {\bf 307}, 841

\bibitem[\protect\astroncite{Couch et~al.}{1998}]{C+98}
Couch, W.~J., Barger, A.~J., Smail, I., Ellis, R.~S., and Sharples, R.~M.:
  1998,
\newblock {\em ApJ} {\bf 497}, 188

\bibitem[\protect\astroncite{Couch et~al.}{1994}]{C+94}
Couch, W.~J., Ellis, R.~S., Sharples, R.~M., and Smail, I.: 1994,
\newblock {\em ApJ} {\bf 430}, 121

\bibitem[\protect\astroncite{{Courteau} et~al.}{1996}]{Courteau96}
{Courteau}, S., {de Jong}, R.~S., and {Broeils}, A.~H.: 1996,
\newblock {\em ApJL} {\bf 457}, L73

\bibitem[\protect\astroncite{{Crampton} et~al.}{1995}]{CFRS5}
{Crampton}, D., {Le Fevre}, O., {Lilly}, S.~J., and {Hammer}, F.: 1995,
\newblock {\em ApJ} {\bf 455}, 96

\bibitem[\protect\astroncite{{de Jong}}{1994}]{deJong94}
{de Jong}, R.~S.: 1994,
\newblock {\em Ph.D. thesis}, Kapteyn Astronomical Inst., (1994)

\bibitem[\protect\astroncite{{de Jong}}{1996}]{deJong96}
{de Jong}, R.~S.: 1996,
\newblock {\em A\&ASS} {\bf 118}, 557

\bibitem[\protect\astroncite{{Diaferio} et~al.}{2001}]{Diaferio}
{Diaferio}, A., {Kauffmann}, G., {Balogh}, M.~L., {White}, S. D.~M., {Schade},
  D., and {Ellingson}, E.: 2001,
\newblock {\em MNRAS} {\bf 323}, 999

\bibitem[\protect\astroncite{Dressler}{1980}]{Dressler}
Dressler, A.: 1980,
\newblock {\em ApJ} {\bf 236}, 351

\bibitem[\protect\astroncite{Dressler et~al.}{1997}]{D+97}
Dressler, A., Oemler, A., Couch, W.~J., Smail, I., Ellis, R.~S., Barger, A.,
  Butcher, H.~R., Poggianti, B.~M., and Sharples, R.~M.: 1997,
\newblock {\em ApJ} {\bf 490}, 577

\bibitem[\protect\astroncite{Dressler et~al.}{1985}]{D+85}
Dressler, A., Thompson, I.~B., and Shectman, S.: 1985,
\newblock {\em ApJ} {\bf 288}, 481

\bibitem[\protect\astroncite{{Ebeling} et~al.}{1998}]{Ebeling98}
{Ebeling}, H., {Edge}, A.~C., {Bohringer}, H., {Allen}, S.~W., {Crawford},
  C.~S., {Fabian}, A.~C., {Voges}, W., and {Huchra}, J.~P.: 1998,
\newblock {\em MNRAS} {\bf 301}, 881

\bibitem[\protect\astroncite{{Ebeling} et~al.}{1996}]{XBACS}
{Ebeling}, H., {Voges}, W., {Bohringer}, H., {Edge}, A.~C., {Huchra}, J.~P.,
  and {Briel}, U.~G.: 1996,
\newblock {\em MNRAS} {\bf 281}, 799

\bibitem[\protect\astroncite{{Fabricant} et~al.}{2000}]{FFD}
{Fabricant}, D., {Franx}, M., and {van Dokkum}, P.: 2000,
\newblock {\em ApJ} {\bf 539}, 577

\bibitem[\protect\astroncite{{Farouki} and {Shapiro}}{1982}]{FS82}
{Farouki}, R.~T. and {Shapiro}, S.~L.: 1982,
\newblock {\em ApJ} {\bf 259}, 103

\bibitem[\protect\astroncite{{Fasano} et~al.}{2000}]{F+00}
{Fasano}, G., {Poggianti}, B.~M., {Couch}, W.~J., {Bettoni}, D., {Kjaergaard},
  P., and {Moles}, M.: 2000,
\newblock {\em ApJ} {\bf 542}, 673

\bibitem[\protect\astroncite{Fukugita et~al.}{1995}]{F+95}
Fukugita, M., Shimasaku, K., and Ichikawa, T.: 1995,
\newblock {\em PASP} {\bf 107}, 945

\bibitem[\protect\astroncite{Giovanelli and Haynes}{1985}]{GH85}
Giovanelli, R. and Haynes, M.: 1985,
\newblock {\em ApJ} {\bf 292}, 404

\bibitem[\protect\astroncite{{Gunn} and {Gott}}{1972}]{GG}
{Gunn}, J.~E. and {Gott}, J. R.~I.: 1972,
\newblock {\em ApJ} {\bf 176}, 1

\bibitem[\protect\astroncite{Hubble and Humason}{1931}]{HH31}
Hubble, E. and Humason, M.~L.: 1931,
\newblock {\em ApJ} {\bf 74}, 43

\bibitem[\protect\astroncite{{Hubble}}{1922}]{Hubble1}
{Hubble}, E.~P.: 1922,
\newblock {\em ApJ} {\bf 56}, 162

\bibitem[\protect\astroncite{{Hubble}}{1926}]{Hubbleseq}
{Hubble}, E.~P.: 1926,
\newblock {\em ApJ} {\bf 64}, 321

\bibitem[\protect\astroncite{{Jones} et~al.}{2000}]{JSC}
{Jones}, L., {Smail}, I., and {Couch}, W.~J.: 2000,
\newblock {\em ApJ} {\bf 528}, 118

\bibitem[\protect\astroncite{{Kauffmann}}{1996}]{K96}
{Kauffmann}, G.: 1996,
\newblock {\em MNRAS} {\bf 281}, 487

\bibitem[\protect\astroncite{{Kauffmann} et~al.}{1993}]{KWG}
{Kauffmann}, G., {White}, S. D.~M., and {Guiderdoni}, B.: 1993,
\newblock {\em MNRAS} {\bf 264}, 201

\bibitem[\protect\astroncite{{Kodama} and {Smail}}{2001}]{KS00}
{Kodama}, T. and {Smail}, I.: 2001,
\newblock {\em MNRAS} {\bf 326}, 637

\bibitem[\protect\astroncite{{Kuntschner} and {Davies}}{1998}]{KD98}
{Kuntschner}, H. and {Davies}, R.~L.: 1998,
\newblock {\em MNRAS} {\bf 295}, L29

\bibitem[\protect\astroncite{{Lacey} and {Cole}}{1994}]{LC94}
{Lacey}, C. and {Cole}, S.: 1994,
\newblock {\em MNRAS} {\bf 271}, 676

\bibitem[\protect\astroncite{Larson et~al.}{1980}]{LTC}
Larson, R.~B., Tinsley, B.~M., and Caldwell, C.~N.: 1980,
\newblock {\em ApJ} {\bf 237}, 692

\bibitem[\protect\astroncite{Lewis et~al.}{1999}]{L+99}
Lewis, A.~D., Ellingson, E., Morris, S.~L., and Carlberg, R.~G.: 1999,
\newblock {\em ApJ} {\bf 517}, 587

\bibitem[\protect\astroncite{{Markevitch}}{1998}]{Markevitch}
{Markevitch}, M.: 1998,
\newblock {\em ApJ} {\bf 504}, 27

\bibitem[\protect\astroncite{{Marleau} and {Simard}}{1998}]{MS}
{Marleau}, F.~R. and {Simard}, L.: 1998,
\newblock {\em ApJ} {\bf 507}, 585

\bibitem[\protect\astroncite{{Metcalfe} et~al.}{2001}]{WHDF5}
{Metcalfe}, N., {Shanks}, T., {Campos}, A., {McCracken}, H.~J., and {Fong}, R.:
  2001,
\newblock {\em MNRAS} {\bf 323}, 795

\bibitem[\protect\astroncite{Metropolis et~al.}{1953}]{Metropolis}
Metropolis, N., Rosenbluth, A., Rosenbluth, M., Teller, A., and Teller, E.:
  1953,
\newblock {\em Journal of Chemical Physics} {\bf 21}, 1087

\bibitem[\protect\astroncite{{Moore} et~al.}{1999}]{harass}
{Moore}, B., {Lake}, G., {Quinn}, T., and {Stadel}, J.: 1999,
\newblock {\em MNRAS} {\bf 304}, 465

\bibitem[\protect\astroncite{Morgan}{1961}]{M61}
Morgan, W.~W.: 1961,
\newblock {\em Proc. Nat. Acad. Sci.} {\bf 47}, 905

\bibitem[\protect\astroncite{{Moss} and {Whittle}}{2000}]{MW00}
{Moss}, C. and {Whittle}, M.: 2000,
\newblock {\em MNRAS} {\bf 317}, 667

\bibitem[\protect\astroncite{{Naim} et~al.}{1997}]{NRG}
{Naim}, A., {Ratnatunga}, K.~U., and {Griffiths}, R.~E.: 1997,
\newblock {\em ApJS} {\bf 111}, 357

\bibitem[\protect\astroncite{{Narayanan} et~al.}{2000}]{NBW}
{Narayanan}, V.~K., {Berlind}, A.~A., and {Weinberg}, D.~H.: 2000,
\newblock {\em ApJ} {\bf 528}, 1

\bibitem[\protect\astroncite{Nulsen}{1982}]{N82}
Nulsen, P. E.~J.: 1982,
\newblock {\em MNRAS} {\bf 198}, 1007

\bibitem[\protect\astroncite{{Odewahn} et~al.}{1996}]{O+96}
{Odewahn}, S.~C., {Windhorst}, R.~A., {Driver}, S.~P., and {Keel}, W.~C.: 1996,
\newblock {\em ApJL} {\bf 472}, L13

\bibitem[\protect\astroncite{Osterbrock}{1960}]{O60}
Osterbrock, D.~E.: 1960,
\newblock {\em ApJ} {\bf 132}, 325

\bibitem[\protect\astroncite{{Ostrander} et~al.}{1998}]{MDS-data}
{Ostrander}, E.~J., {Nichol}, R.~C., {Ratnatunga}, K.~U., and {Griffiths},
  R.~E.: 1998,
\newblock {\em AJ} {\bf 116}, 2644

\bibitem[\protect\astroncite{Poggianti et~al.}{1999}]{P+99}
Poggianti, B.~M., Smail, I., Dressler, A., Couch, W.~J., Barger, A.~J.,
  Butcher, H., Ellis, R.~S., and Oemler, A.: 1999,
\newblock {\em ApJ} {\bf 518}, 576

\bibitem[\protect\astroncite{Postman and Geller}{1984}]{PG84}
Postman, M. and Geller, M.~J.: 1984,
\newblock {\em ApJ} {\bf 281}, 95

\bibitem[\protect\astroncite{{Ratnatunga} et~al.}{1999}]{MDS-morph3}
{Ratnatunga}, K.~U., {Griffiths}, R.~E., and {Ostrander}, E.~J.: 1999,
\newblock {\em AJ} {\bf 118}, 86

\bibitem[\protect\astroncite{{Roche} et~al.}{1996}]{MDS-numcts}
{Roche}, N., {Ratnatunga}, K., {Griffiths}, R.~E., {Im}, M., and {Neuschaefer},
  L.: 1996,
\newblock {\em MNRAS} {\bf 282}, 1247

\bibitem[\protect\astroncite{{Roos} and {Norman}}{1979}]{RN}
{Roos}, N. and {Norman}, C.~A.: 1979,
\newblock {\em A\&A} {\bf 76}, 75

\bibitem[\protect\astroncite{{Saglia} et~al.}{2000}]{S+00}
{Saglia}, R.~P., {Maraston}, C., {Greggio}, L., {Bender}, R., and {Ziegler},
  B.: 2000,
\newblock {\em A\&A} {\bf 360}, 911

\bibitem[\protect\astroncite{{Schade} et~al.}{1997}]{Schade97}
{Schade}, D., {Barrientos}, L.~F., and {Lopez-Cruz}, O.: 1997,
\newblock {\em ApJL} {\bf 477}, L17

\bibitem[\protect\astroncite{Schade et~al.}{1996a}]{S2+96}
Schade, D., Carlberg, R.~G., Yee, H. K.~C., L\'{o}pez-Cruz, O., and Ellingson,
  E.: 1996a,
\newblock {\em ApJL} {\bf 464}, 103

\bibitem[\protect\astroncite{Schade et~al.}{1995}]{Schade95}
Schade, D., Lilly, S.~J., Crampton, D., LeF\`{e}vre, O., Hammer, F., and
  Tresse, L.: 1995,
\newblock {\em ApJ} {\bf 451}, 1

\bibitem[\protect\astroncite{Schade et~al.}{1996b}]{S3+96}
Schade, D., Lilly, S.~J., LeF\`{e}vre, O., Hammer, F., and Crampton, D.: 1996b,
\newblock {\em ApJ} {\bf 464}, 79

\bibitem[\protect\astroncite{Simard et~al.}{2001}]{Gim2d}
Simard, L., Willmer, C. N.~A., Vogt, N.~P., Sarajedini, V.~L., Phillips, A.~C.,
  Koo, D.~C., Im, M., Illingworth, G.~D., Gebhardt, K., and Faber, S.~M.: 2001,
\newblock {\em in preparation}

\bibitem[\protect\astroncite{{Simien} and {de Vaucouleurs}}{1986}]{SdV}
{Simien}, F. and {de Vaucouleurs}, G.: 1986,
\newblock {\em ApJ} {\bf 302}, 564

\bibitem[\protect\astroncite{Smail et~al.}{1997}]{Smail}
Smail, I., Dressler, A., Couch, W.~J., Ellis, R.~S., Oemler, A., Butcher, H.,
  and Sharples, R.~M.: 1997,
\newblock {\em ApJS} {\bf 110}, 213

\bibitem[\protect\astroncite{{Smail} et~al.}{2001}]{Smail01}
{Smail}, I., {Kuntschner}, H., {Kodama}, T., {Smith}, G.~P., {Packham}, C.,
  {Fruchter}, A.~S., and {Hook}, R.~N.: 2001,
\newblock {\em MNRAS} {\bf 323}, 839

\bibitem[\protect\astroncite{{Smith} et~al.}{2001a}]{GPS1}
{Smith}, G.~P., {Kneib}, J., {Ebeling}, H., {Czoske}, O., and {Smail}, I.:
  2001a,
\newblock {\em ApJ} {\bf 552}, 493

\bibitem[\protect\astroncite{{Smith} et~al.}{2001b}]{GPS2}
{Smith}, G.~P., Smail, I., Kneib, J., Czoske, O., Ebeling, H., Edge, A., Pello,
  R., Ivison, R.~J., Packham, C., and {Le Borgne}, J.: 2001b,
\newblock {\em MNRAS} p. submitted

\bibitem[\protect\astroncite{{Solanes} et~al.}{2001}]{Solanes01}
{Solanes}, J.~., {Manrique}, A., {Garc{\' i}a-G{\' o}mez}, C., {Gonz{\'
  a}lez-Casado}, G., {Giovanelli}, R., and {Haynes}, M.~P.: 2001,
\newblock {\em ApJ} {\bf 548}, 97

\bibitem[\protect\astroncite{{Solanes} and {Salvador-Sole}}{1992}]{SSS}
{Solanes}, J.~M. and {Salvador-Sole}, E.: 1992,
\newblock {\em ApJ} {\bf 395}, 91

\bibitem[\protect\astroncite{{Somerville} and {Primack}}{1999}]{SP99}
{Somerville}, R.~S. and {Primack}, J.~R.: 1999,
\newblock {\em MNRAS} {\bf 310}, 1087

\bibitem[\protect\astroncite{Spitzer and Baade}{1951}]{SB51}
Spitzer, L. and Baade, W.: 1951,
\newblock {\em ApJ} {\bf 113}, 413

\bibitem[\protect\astroncite{{Struble} and {Rood}}{1999}]{SR99}
{Struble}, M.~F. and {Rood}, H.~J.: 1999,
\newblock {\em ApJS} {\bf 125}, 35

\bibitem[\protect\astroncite{{Tran} et~al.}{2001}]{Tran}
{Tran}, K.~H., {Simard}, L., {Zabludoff}, A.~I., and {Mulchaey}, J.~S.: 2001,
\newblock {\em ApJ} {\bf 549}, 172

\bibitem[\protect\astroncite{van~den Bergh}{1991}]{vdB91}
van~den Bergh, S.: 1991,
\newblock {\em PASP} {\bf 103}, 390

\bibitem[\protect\astroncite{{van den Bergh}}{1997}]{vdB97}
{van den Bergh}, S.: 1997,
\newblock {\em AJ} {\bf 113}, 2054

\bibitem[\protect\astroncite{{van Dokkum} et~al.}{1998}]{vd98}
{van Dokkum}, P.~G., {Franx}, M., {Kelson}, D.~D., {Illingworth}, G.~D.,
  {Fisher}, D., and {Fabricant}, D.: 1998,
\newblock {\em ApJ} {\bf 500}, 714

\bibitem[\protect\astroncite{Vikhlinin et~al.}{1998}]{V+98}
Vikhlinin, A., McNamara, B., Forman, W., Jones, C., and Quintana, H.: 1998,
\newblock {\em ApJL} {\bf 498}, 21

\bibitem[\protect\astroncite{{Vollmer} et~al.}{2001}]{V+01}
{Vollmer}, B., {Cayatte}, V., {van Driel}, W., {Henning}, P.~A.,
  {Kraan-Korteweg}, R.~C., {Balkowski}, C., {Woudt}, P.~A., and {Duschl},
  W.~J.: 2001,
\newblock {\em A\&A} {\bf 369}, 432

\bibitem[\protect\astroncite{White and Sarazin}{1991}]{WS91}
White, R.~E. and Sarazin, C.~L.: 1991,
\newblock {\em ApJ} {\bf 367}, 476

\bibitem[\protect\astroncite{Whitmore and Gilmore}{1991}]{WG}
Whitmore, B.~C. and Gilmore, D.~M.: 1991,
\newblock {\em ApJ} {\bf 367}, 64

\bibitem[\protect\astroncite{Whitmore et~al.}{1993}]{WGJ}
Whitmore, B.~C., Gilmore, D.~M., and Jones, C.: 1993,
\newblock {\em ApJ} {\bf 407}, 489

\bibitem[\protect\astroncite{{Wu} et~al.}{1999}]{WXF}
{Wu}, X., {Xue}, Y., and {Fang}, L.: 1999,
\newblock {\em ApJ} {\bf 524}, 22

\bibitem[\protect\astroncite{{Xue} and {Wu}}{2000}]{XW}
{Xue}, Y. and {Wu}, X.: 2000,
\newblock {\em ApJ} {\bf 538}, 65

\bibitem[\protect\astroncite{{Yoshizawa} and {Wakamatsu}}{1975}]{YW75}
{Yoshizawa}, M. and {Wakamatsu}, K.: 1975,
\newblock {\em A\&A} {\bf 44}, 363

\bibitem[\protect\astroncite{{Zabludoff} and {Mulchaey}}{1998}]{ZM98}
{Zabludoff}, A.~I. and {Mulchaey}, J.~S.: 1998,
\newblock {\em ApJ} {\bf 496}, 39

\bibitem[\protect\astroncite{{Zabludoff} and {Mulchaey}}{2000}]{ZM00}
{Zabludoff}, A.~I. and {Mulchaey}, J.~S.: 2000,
\newblock {\em ApJ} {\bf 539}, 136

\end{thebibliography}
\end{document}